\documentclass[12pt]{article}
\usepackage[english]{babel}
\usepackage{graphicx,subfigure}
\usepackage{amssymb,amsthm,amsmath,amsfonts}
\usepackage{float}
\usepackage[colorlinks=true,linkcolor=blue,citecolor=red, urlcolor=green]{hyperref} 
\usepackage{cite}

\begin{document}

\title{Fermionic operatorial model of a system with competitive and cooperative interactions}

\author{M.~Gorgone, G.~Inferrera, F.~Oliveri\\
\ \\
{\footnotesize Department of Mathematical and Computer Sciences,}\\
{\footnotesize Physical Sciences and Earth Sciences, University of Messina}\\
{\footnotesize Viale F. Stagno d'Alcontres 31, 98166 Messina, Italy}\\
{\footnotesize mgorgone@unime.it; guinferrera@unime.it; foliveri@unime.it}
}

\date{Published on \emph{ Int. J. Theor. Phys. \textbf{62}, 241 (2023).}}

\maketitle

\begin{abstract}
An operatorial model of a system made by $N$ agents interacting each other with mechanisms that can be thought of as cooperative or competitive is presented. We associate to each agent an annihilation, creation and number fermionic operator, and interpret the mean values of the number operators over an initial condition as measures of the agents' wealth status. The dynamics of the system is assumed to be ruled by a Hermitian Hamiltonian operator 
$\mathcal{H}$, and the classical Heisenberg view  is used. The dynamical outcome is then enriched by using the recently introduced variant of $(\mathcal{H},\rho)$--induced dynamics, where $\rho$ denotes a rule that periodically modifies some of the parameters involved in $\mathcal{H}$. The agents are partitioned in 
three subgroups, one interacting each other only with a competitive mechanism, one interacting each other only with a 
cooperative mechanism, and one opportunist subgroup able to compete and cooperate. Some numerical simulations show that the $(\mathcal{H},\rho)$--induced dynamics approach makes, in all the cases, the cooperative subgroup definitely to be more efficient in improving its wealth status than the other subgroups. 
\end{abstract}

$(\mathcal{H},\rho)$--induced dynamics; Competitive, cooperative and opportunist agents.


\section{Introduction}
\label{sec:intro}

In the last decade, ladder operators typical of quantum mechanics \cite{Roman,Merzbacher} have 
been  used in the mathematical modeling of several kinds of classical macroscopic systems (see 
\cite{bagbook1,bagbook2,fffbook}).
In fact, many recent contributions in the area of quantum--like modeling outside physics, 
using the formalism of second quantization, and in particular the number representation, revealed 
successful in different contexts: models of stock markets 
\cite{
Bagarello-stock3,Bagarello-stock4}, 
social science and decision--making processes \cite{QSC,qdm1,qdm2,qdm3,qdm4}, 
population migration and crowd dynamics \cite{BO_migration,Gargano-populations,BGO_crowds}, 
bio--ecological modelling 
\cite{BO_ecomod,BCO_desert,DSO_RM2016,BDSGO_GoL}, political systems   
\cite{pol1,all1,all2,PKEH2016,BG17,DSO_turncoat2017,DSGO_turncoat2017,DSGO_opinion2017},  information spreading in a network \cite{BGO-information}. The motivations for using quantum formalism to describe classical 
situations have been widely discussed (see \cite{bagbook1,bagbook2,fffbook,QSC,qdm1,qdm2}, and references 
therein). 

In an operatorial model of a macroscopic system $\mathcal{S}$, the unknowns  are operators living 
in a Hilbert space $\mathbb{H}$ that can be finite or infinite dimensional (depending on the 
choice of using the fermionic rather 
than the bosonic representation). What we need is to identify the observables of  $\mathcal{S}$, 
\emph{i.e.}, the self--adjoint operators relevant for the description of the system itself, and 
compute the mean values of such operators evaluated on the state corresponding to an assigned 
initial condition; so doing we obtain some real valued functions that can be phenomenologically associated 
to some macroscopic quantities. 

The time evolution of such a system is recovered by assuming the dynamics ruled by a self--adjoint 
time--independent Hamiltonian operator $\mathcal{H}$ embedding the various interactions occurring 
within the actors of the system. In particular, to take the computational complexity low, in the 
following we will limit ourselves to a quadratic Hamiltonian. 
Using Heisenberg view, the time evolution of an observable operator
$X$ of the macroscopic system $\mathcal{S}$ under consideration is given by 
$X(t)=\exp(\textrm{i}\mathcal{H}t)X\exp(-\textrm{i}\mathcal{H}t)$. 

There is a cost to pay in using a time--independent 
self--adjoint  quadratic Hamiltonian operator: in fact, we obtain time evolutions at most 
quasiperiodic. 
Then, if the system $\mathcal{S}$ we wish to model has some asymptotic \emph{final state}, 
it is clear that such a description cannot be effective, and the framework needs to be enriched, 
if not completely changed. This can be done, for instance, considering Hamiltonian operators 
involving terms more than quadratic, or inserting an explicit time dependence in the coefficients 
of the Hamiltonian itself, or admitting that the actors of the system interact with some infinite 
reservoir, or using non--Hermitian Hamiltonians \cite{Bagarello-MPAG}. However, in all these cases, many technical and computational difficulties arise.

In some recent papers 
\cite{BDSGO_GoL,DSO_turncoat2017}, an approach, 
named $(\mathcal{H},\rho)$--induced dynamics, revealed successful in enriching the dynamical 
outcome still using a quadratic self--adjoint time--independent Hamiltonian. 

With such an approach, we take into account effects which may occur during the time evolution of 
the system, and which can not be easily included in a purely Hamiltonian description. In 
particular, during the evolution of the system, driven by a time--independent Hermitian 
Hamiltonian $\mathcal{H}$, at fixed times some checks on the system are performed, and these 
periodical measures on the system are used to change the state of the system itself according to 
an explicit prescription \cite{BDSGO_GoL}, or some of the parameters entering the Hamiltonian 
itself, without modifying 
the functional form of the Hamiltonian itself (see \cite{BDSGO-PhysicaA}, where the authors analyzed in general this approach, and provided a comparison with the well established frameworks where the Hamiltonian operator is 
time--dependent or where the system is open and interacting with a reservoir). 
This approach proved to be quite efficient in 
operatorial models of stressed bacterial populations \cite{DSO_RM2016}, as well as in models of 
political systems affected by turncoat--like behaviors of part of their members,  \emph{i.e.}, 
systems characterized by internal fluxes between different political parties 
\cite{DSO_turncoat2017,DSGO_turncoat2017,DSGO_opinion2017}. 
In some sense, such an approach allows 
us to describe a sort of discrete self--adaptation of the model depending on the evolution of the 
state of the system. The strategy of the $(\mathcal{H},\rho)$--induced dynamics may give 
interesting results if the rule $\rho$ is not introduced as a mere mathematical expedient, but  
is somehow physically justified. 

In this paper, we implement a model for a system $\mathcal{S}$ 
made by a finite number of agents interacting each other with competitive as well as cooperative mechanisms. We associate to each agent ${A}_j$ an annihilation $a_j$ and a creation $a_j^\dagger$  fermionic operator; the mean value associated to the number operator $\widehat{n}_j=a_j^\dagger a_j$ is interpreted as a measure of the wealth of 
${A}_j$. The Hamiltonian ruling the dynamics of this system involves terms whose meaning can be  assimilated to competitive or cooperative interactions. 
As an illustrative example, we first consider a simple model made by seven agents divided 
in three subgroups (three agents with competitive interactions, three agents with cooperative 
interactions, and one opportunist agent with both types of interaction); by choosing suitable values 
to the interaction parameters, we present the results of some numerical simulations. The dynamical 
outcomes of classical Heisenberg dynamics, as well as that of $(\mathcal{H},\rho)$--induced 
dynamics approach  are given. Moreover, a more complex model where the agents are spatially 
distributed in a one--dimensional torus, with interaction parameters depending on the distance 
between the agents, is considered.

The structure of the paper is the following. 
In Section~\ref{sec:model},  to keep the paper almost self--contained, we briefly review few 
useful notions about quantum mechanics and the number representation for fermions. Then, we 
present our general model, whose time evolution is ruled by a time--independent self--adjoint 
quadratic Hamiltonian operator: competition and cooperation effects are taken into account. 
Because of the quadratic expression of the Hamiltonian, the Heisenberg equations turn out to be 
linear, so that, at least formally, we can deduce analytically the solution. 
Subsection~\ref{sec:rule} briefly describes the $(\mathcal{H},\rho)$--induced dynamics approach; 
the  physical meaning of the \emph{rule} we consider is also clarified. 
In Section~\ref{sec:7modemodel},  we present a simple model made by seven agents and perform some 
numerical simulations. A more sophisticated model, involving one hundred  
agents spatially distributed in a one--dimensional torus, with nonhomogeneous interaction parameters, is analyzed in Section~\ref{sec:manymodemodel}. Finally, Section~\ref{sec:conclusions} 
contains some concluding remarks. 

\section{The fermionic operatorial model}
\label{sec:model}
In this Section, we briefly sketch the basic notions of fermionic operatorial formalism. The 
operatorial model we are interested to aims to describe a system $\mathcal{S}$ made by $N$ 
interacting agents $A_1,\ldots, A_N$; an annihilation ($a_j$), a creation ($a^\dagger_j$), and an 
occupation number ($\widehat n_j=a^\dagger_j a_j$) fermionic operator is associated to each agent. 
Their interactions are ruled by a self--adjoint 
time--independent Hamiltonian operator. Adopting the Heisenberg viewpoint, we derive the 
differential equations ruling the dynamics.

According to the formalism of second quantization, fermionic operators satisfy the \emph{canonical 
anticommutation relations} 
\begin{equation}
\label{CAR}
\{a_j, a_k\}=0,\quad \{a_j^\dagger, a_k^\dagger\}=0,\quad
\{a_j, a_k^\dagger\}=\delta_{j,k}\mathbb{I},
\end{equation}
$j,k=1,\ldots, N$, $\mathbb{I}$ being the identity operator, and $\{u,v\}=uv+vu$ the 
anticommutator between $u$ and $v$. 
The Hilbert space $\mathbb{H}$ in which the fermionic operators  
are defined is constructed as the linear span of the orthonormal set of vectors
\begin{equation}
\label{fermion_vectors}
\varphi_{n_1,n_2.\ldots,n_N}=
(a_1^\dagger)^{n_1}(a_2^\dagger)^{n_2}\cdots(a_N^\dagger)^{n_N}\varphi_{0,0,\ldots,0}, 
\end{equation} 
generated by acting on the \emph{vacuum} $\varphi_{0,0,\ldots,0}$ (\emph{i.e.}, an eigenvector of 
all the annihilation operators) with the operators 
$(a_\ell^\dagger)^{n_\ell}$, $n_\ell=0,1$ for $\ell=1,\ldots,N$; therefore, it is 
$\dim(\mathbb{H})=2^N$.  
The vector $\varphi_{n_1,n_2,\ldots,n_N}$ means that to the $k$--th agent it is initially assigned 
a mean value equal to $n_k$ ($k=1,\ldots,N$). We have 
\begin{equation} 
\widehat n_k\varphi_{n_1,n_2,\ldots,n_N}=n_k\varphi_{n_1,n_2,\ldots,n_N},\qquad k=1,\ldots,N. 
\end{equation}
The interpretation we give to the mean values $n_k$ ($k=1,\ldots,N$) is that of a measure of the 
wealth of the $k$--th agent. The interactions we consider to occur among the agents in the system 
$\mathcal{S}$ fall in two different classes: competition and cooperation.

Let the dynamics of $\mathcal{S}$ be governed by the self--adjoint Hamiltonian
\begin{equation}
\mathcal{H}=\mathcal{H}_0+ \mathcal{H}_I,
\label{hamiltonian}
\end{equation}
with
\begin{equation}
\left\{
\begin{aligned}
&\mathcal{H}_0 = \sum_{k=1}^N\omega_k a_k^\dagger a_k,\\
&\mathcal{H}_I=
\sum_{1\le j<k \le N}\lambda_{j,k}\left(a_j\,a_k^\dagger+a_k\,a_j^\dagger\right)
+\sum_{1\le j<k \le N}\mu_{j,k}\left(a_j^\dagger\,a_k^\dagger+a_k\,a_j\right),
\end{aligned}
\right.
\end{equation}
where the constants $\omega_j$, $\lambda_{j,k}$ and $\mu_{j,k}$ are real positive quantities; we 
remark that in concrete applications not all parameters $\lambda_{j,k}$ and $\mu_{j,k}$ have to be 
non--vanishing. 

The contribution $\mathcal{H}_0$ is the free part of the Hamiltonian, and $\omega_{k}$ are 
parameters somehow related to the inertia of the operators associated to the agents of 
$\mathcal{S}$: in fact, they can be thought of as a measure of the tendency of each degree of 
freedom to stay constant in time \cite{bagbook1}. 

On the contrary, $\mathcal{H}_I$ rules the interactions among the agents. These interactions split 
in two contributions:
\begin{itemize}
\item the term $\lambda_{j,k}\left(a_j\,a_k^\dagger+a_k\,a_j^\dagger\right)$ can be interpreted as 
a competitive  contribution, and the coefficient $\lambda_{j,k}$  gives a measure of the strength 
of the interaction between the agents $A_j$ and $A_k$; more precisely, the contribution 
$a_{j}a_{k}^\dagger$ is a competition term since it destroys a \emph{particle} for the agent 
associated to $a_{j}$ and creates a \emph{particle} for the agent associated to $a_{k}$; the 
adjoint term $a_{k}a_{j}^\dagger$ swaps the roles of the two agents; in other words, \emph{the 
loss (gain) of an agent is the gain (loss) of the other agent};
\item the term $\mu_{j,k}\left(a_j^\dagger\,a_k^\dagger+a_k\,a_j\right)$ can be interpreted as a 
cooperative contribution, and $\mu_{j,k}$ is a measure of the strength of this cooperation;
the term $a_j^\dagger a_k^\dagger$ creates a \emph{particle} for both agents, and the adjoint part 
destroys a \emph{particle} for both agents: in other words, \emph{the gain (loss) of an agent is 
the gain (loss) of the other agent}. 
\end{itemize}

Adopting the Heisenberg view for the dynamics, the time evolution of the annihilation operators 
$a_j(t)$ is ruled by 
\begin{equation}
\frac{d a_j}{dt} =\textrm{i} \left[H,a_j\right] ,\qquad j=1,\ldots,N,
\end{equation}
$[\mathcal{H},a_j]=\mathcal{H}a_j-a_j\mathcal{H}$ being the commutator between $\mathcal{H}$ 
and $a_j$. Thus, we have a system of linear ordinary differential equations (whose unknowns are operators) \begin{equation}
\label{eq-annihilation}
\frac{d a_j}{dt}=\textrm{i}\left(-\omega_j a_j + 
\sum_{1\le \ell < j}\left(\lambda_{\ell,j} a_\ell + \mu_{\ell,j} a_\ell^\dagger\right)+
\sum_{j< k \le N }\left(\lambda_{j,k} a_k -\mu_{j,k} a_k^\dagger \right)\right),
\end{equation}
that have to be solved with suitable initial conditions $a_j(0)=a_j^0$, $j=1,\ldots,N$. Because 
each operator $a_j$ is a square matrix of order $2^N$, in principle we have to solve a system of $N\cdot 4^N$ linear 
ordinary differential equations.

Nevertheless, because of the linearity, we can adopt a \emph{reduced} approach. Let us introduce a 
formal vector
$\mathbf{A}\equiv\left(a_1,\ldots,a_N,a_1^\dagger,\ldots, a_N^\dagger\right)^T$ 
(the superscript ${}^T$ stands for transposition) and the square matrix of order $2N$ 
\[
\Gamma= \left[
\begin{array}{cc}
\Gamma_0 & \Gamma_1 \\
-\Gamma_1 & -\Gamma_0 
\end{array}
\right],
\]
where the $N\times N$ symmetric block $\Gamma_0$ and the antisymmetric block $\Gamma_1$ are
\[
\Gamma_0 = \left[
\begin{array}{ccccc}
-\omega_1 & \lambda_{1,2} & \cdots & \cdots & \lambda_{1,N} \\
\lambda_{1,2} & -\omega_2 & \lambda_{2,3} & \cdots & \lambda_{2,N}  \\
\cdots & \cdots & \cdots & \cdots & \cdots \\
\cdots & \cdots & \cdots & \cdots & \cdots \\
\lambda_{1,N} & \lambda_{2,N} & \cdots & \lambda_{N-1,N} & -\omega_N
\end{array}
\right]
\]  
and
\[
\Gamma_1 = \left[
\begin{array}{ccccc}
0 & -\mu_{1,2} & \cdots & \cdots & -\mu_{1,N} \\
\mu_{1,2} & 0 & -\mu_{2,3} & \cdots & -\mu_{2,N}  \\
\cdots & \cdots & \cdots & \cdots & \cdots \\
\cdots & \cdots & \cdots & \cdots & \cdots \\
\mu_{1,N} & \mu_{2,N} & \cdots & \mu_{N-1,N} & 0
\end{array}
\right],
\]  
respectively.

With these positions, equations~(\ref{eq-annihilation}), together with their adjoint version, 
write in the compact form
\[
\frac{d\mathbf{A}}{dt}=\textrm{i}\Gamma \mathbf{A},\qquad \mathbf{A}(0)=\mathbf{A}^0,
\]
whose solution formally is
\[
\mathbf{A}(t)= \mathcal{B}(t)\,\mathbf{A}^0, \qquad \mathcal{B}(t)=
\exp\left(\textrm{i}\Gamma t\right).
\] 
Now, let us define the vector 
\[
\Phi=\sqrt{n_1^0}\varphi_{1,0,\ldots,0}+\sqrt{n_2^0}\varphi_{0,1,\ldots,0}+\cdots+\sqrt{n_N^0}\varphi_{0,0,\ldots,1},
\]
where $(n_1^0,n_2^0, \ldots, n_N^0)$, such that $n_1^0+\cdots+n_N^0=1$, whence $\Phi$ is unitary, represent the initial values of the mean values of the number 
operators associated to the agents of the system.  
If $B_{j,k}$ is the generic entry of matrix $\mathcal{B}(t)$, we have
\begin{equation}
\begin{aligned}
&a^\dagger_k(t)=\sum_{j=1}^{N}\left(B_{k+N,j}a_j^0+B_{k+N,j+N}{a_j^0}^\dagger\right),\\
&a_k(t)=\sum_{j=1}^{N}\left(B_{k,j}a_j^0+B_{k,j+N}{a_j^0}^\dagger\right),
\end{aligned}
\end{equation}
whereupon the formula
\begin{equation}
\label{observable}
n_k(t)=\left\langle\Phi,a^\dagger_k(t)a_k(t)\Phi\right\rangle,
\end{equation}
using the canonical anticommutation relations (\ref{CAR}), provides
the mean values of the number operators at time $t$:
\begin{equation}
\begin{aligned}
n_{k}(t)&=\sum_{i=1}^{N}\Phi_{i}^2\sum_{\ell=1}^{N}B_{k,f(\ell,k)}B_{k+N,g(\ell,k)}\\
&+\sum_{i=1}^{N-1}\sum_{j=i+1}^{N}\Phi_i\Phi_j\left(B_{k,i}B_{k+N,j+N}+B_{k,j}B_{k+N,i+N}
\right.\\
&\left.\qquad-B_{k,i+N}B_{k+N,j}-B_{k,j+N}B_{k+N,i}\right),
\end{aligned}
\end{equation}
where
\[
f(\ell,k)=\left\{
\begin{array}{lll}
k \quad & \hbox{if}\quad &k=\ell,\\
k+N\quad &\hbox{if}\quad & k\ne\ell,
\end{array}
\right.
\qquad g(\ell,k)=\left\{
\begin{array}{lll}
k+N \quad & \hbox{if}\quad & k=\ell,\\
k\quad & \hbox{if}\quad & k\ne\ell .
\end{array}
\right. 
\]
The real functions given by (\ref{observable}) are represented as measures of the wealth of the 
agents of the system. Due to the quadratic form of  the Hamiltonian $H$, it is not a surprise that the solution exhibits 
a never ending oscillatory behavior. 

\subsection{$(\mathcal{H},\rho)$--induced dynamics}
\label{sec:rule}
According to the approach named  $(\mathcal{H},\rho)$--induced dynamics, we modify the standard 
Heisenberg dynamics by introducing some \emph{rules} able to account for some effects that cannot 
be embedded in the definition of the Hamiltonian, unless we do not assume explicitly a time--dependent Hamiltonian or consider an open quantum system including a reservoir, with consequent 
increase of technical difficulties (see \cite{BDSGO-PhysicaA}). 

The  \emph{rule} we consider is nothing more than a law that modifies periodically some of the 
values of the parameters involved in the Hamiltonian as a consequence of the evolution of the 
system. The underlying idea is that the model adjusts itself during the time evolution;  since the 
model involves some actors, the modifications of some of the parameters entering the Hamiltonian 
reflect some changes in the intensity of the interactions according to the evolution of their 
state. In other words,  these modifications may be thought of as a surreptitious way to take into 
account the influence of the external world, even if this action is induced by the evolution of 
the system itself; actually, the evolution of the state of the system does influence the attitudes 
of the different agents!

Hereafter, we briefly sketch how the procedure works (see \cite{BDSGO-PhysicaA}, and references 
therein, for further details). 
Let us start considering a self--adjoint time--independent quadratic Hamiltonian operator 
$\mathcal{H}^{(1)}$; according to Heisenberg view, we can compute, in a time interval of 
length $\tau>0$, the evolution of annihilation and creation operators, whereupon, choosing an 
initial condition for the mean values of the number operators, obtain their time evolution (our 
observables). According to the values of the observables at time $\tau$, or to their 
variations in the time interval $[0,\tau]$,  we modify some of the parameters involved 
in $\mathcal{H}^{(1)}$. In this way, we get a new 
Hamiltonian operator $\mathcal{H}^{(2)}$, having the same functional form as 
$\mathcal{H}^{(1)}$, but (in general) 
with different values of (some of) the involved parameters, and follow the continuous evolution of 
the system under the action of this new Hamiltonian for the next time interval of length $\tau$. 
Actually, we do not restart the evolution of the system from a new initial condition, but simply 
continue to follow the evolution with the only difference that for $t\in]\tau,2\tau]$ a new 
Hamiltonian $\mathcal{H}^{(2)}$ rules the process. And so on. 

From a mathematical point of view,  the rule is just a map from 
$\mathbb{R}^p$ into $\mathbb{R}^p$ acting on the space of the $p$ parameters involved in the 
Hamiltonian. The global evolution is governed by a sequence of similar Hamiltonian operators, and 
the parameters entering the model can be considered stepwise (in time) constant.

More specifically, let us consider a time interval $[0,T]$ where we follow the evolution of the 
system, and split it in $n = T/\tau$ subintervals of length $\tau$. Assume for simplicity $n$ to 
be integer. In the $k$--th subinterval $](k-1)\tau,k\tau]$ consider an Hermitian Hamiltonian 
$H^{(k)}$ ruling the dynamics and apply Heisenberg view. Therefore, the global dynamics comes from the sequence of Hamiltonians
\begin{equation}
\mathcal{H}^{(1)} \xrightarrow{\tau} \mathcal{H}^{(2)} \xrightarrow{\tau}\mathcal{H}^{(3)} \xrightarrow{\tau} \cdots \xrightarrow{\tau}\mathcal{H}^{(n)},
\end{equation}
and the complete evolution in the interval $[0,T]$ is obtained by glueing the local evolutions
in each subinterval.

This kind of rule--induced stepwise dynamics clearly may generate discontinuities in the first 
order derivatives of the operators, but prevents the occurrence of jumps in their evolutions and, 
consequently, in the mean values of the number operators. By adopting this rule, we are implicitly 
considering the possibility of having a time--dependent Hamiltonian. However, the time dependence 
is, in our case, of a very special form: in each interval $](k-1)\tau, k\tau]$ the Hamiltonian 
does not depend on time, but in $k\tau$ some changes may occur, according to how the system is 
evolving. For this reason, our Hamiltonian can be considered piecewise constant in time. A 
comparison of this approach with that related to an explicitly time--dependent Hamiltonian is 
discussed in \cite{BDSGO-PhysicaA}.

\section{A rather simple model}
\label{sec:7modemodel}
\begin{figure}
\begin{center}
\includegraphics[width=0.65\textwidth]{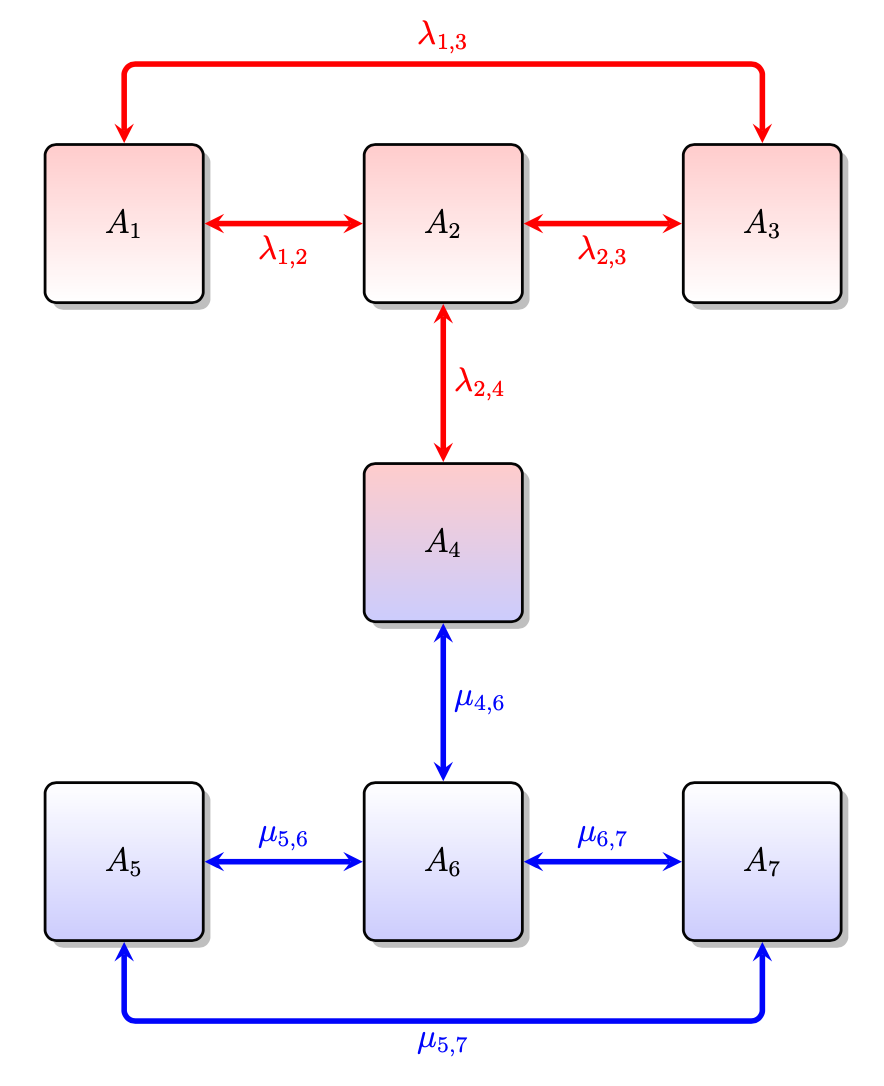}
\caption{\label{fig:7modes}Schematic view of a system with seven agents.}
\end{center}
\end{figure}
As a first simple example to illustrate the above considerations, let us consider a system 
$\mathcal{S}$ composed by seven agents interacting according to the 
scheme shown in Figure~\ref{fig:7modes}. In the system $\mathcal{S}$ we can recognize: 
\begin{itemize}
\item a purely competitive subsystem $\mathcal{S}_1$, made by the agents $A_1$, $A_2$, $A_3$, 
interacting each other in a competitive way; 
\item an \emph{opportunist} subsystem $\mathcal{S}_2$, made by the agent $A_4$, competing with 
$A_2$ and cooperating with $A_6$;
\item a purely cooperative subsystem  $\mathcal{S}_3$, made by the agents $A_5$, $A_6$, $A_7$, 
interacting each other in a cooperative way.
\end{itemize}
\begin{figure}
\begin{center}
\subfigure[]{\includegraphics[width=0.47\textwidth]{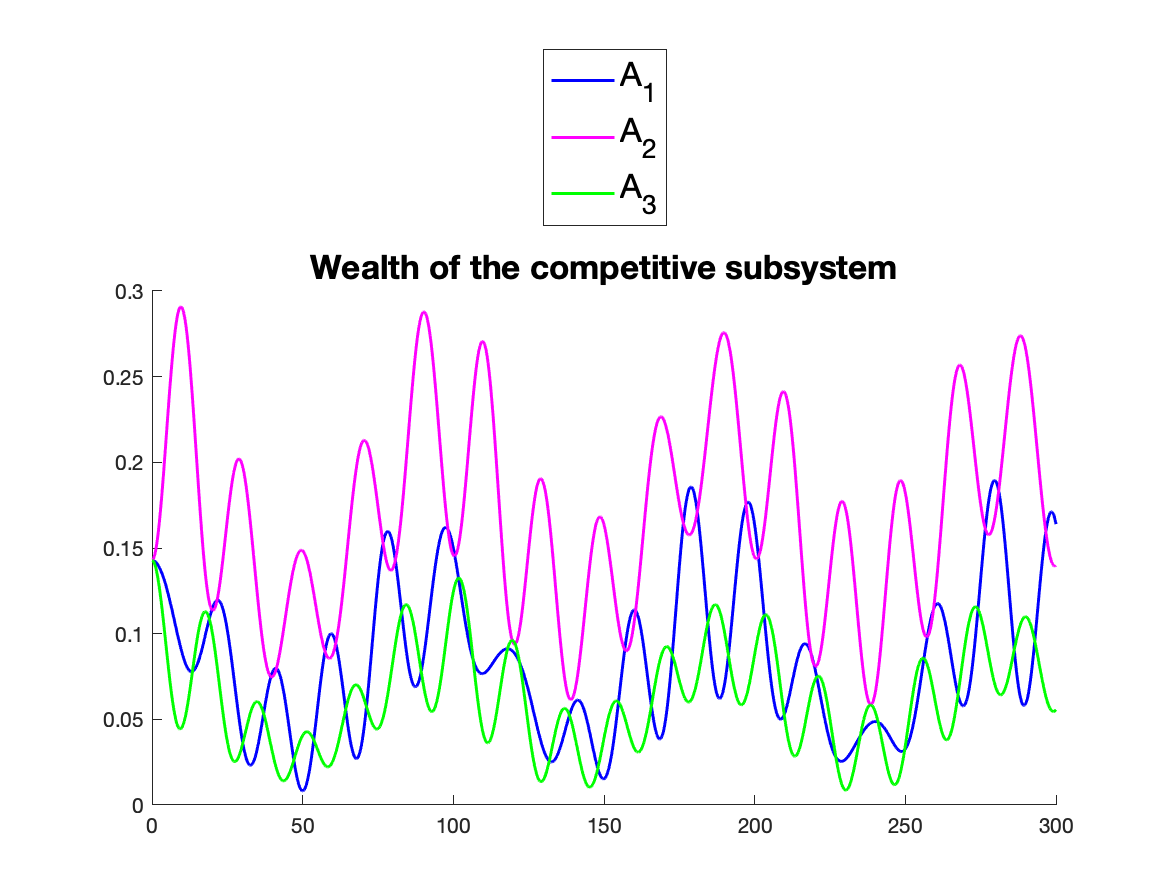}}
\subfigure[]{\includegraphics[width=0.47\textwidth]{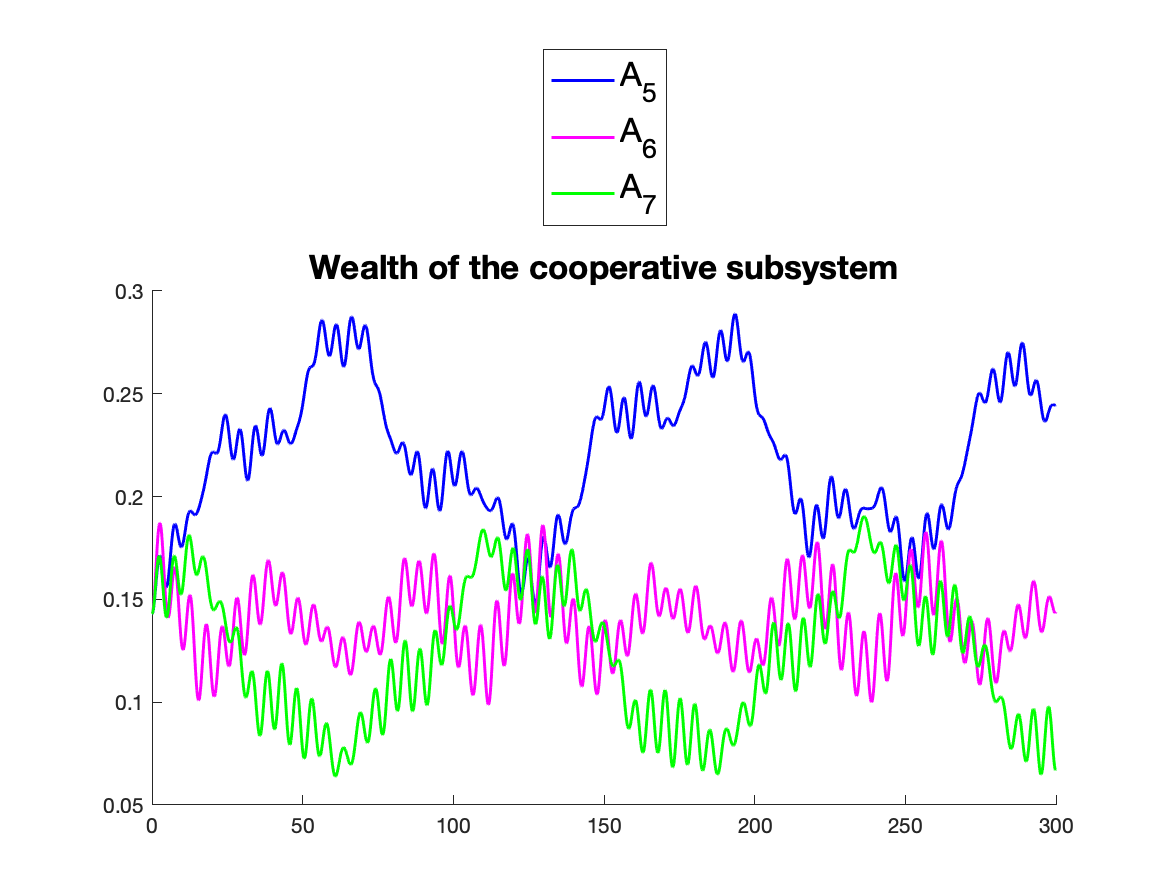}}\\
\subfigure[]{\includegraphics[width=0.47\textwidth]{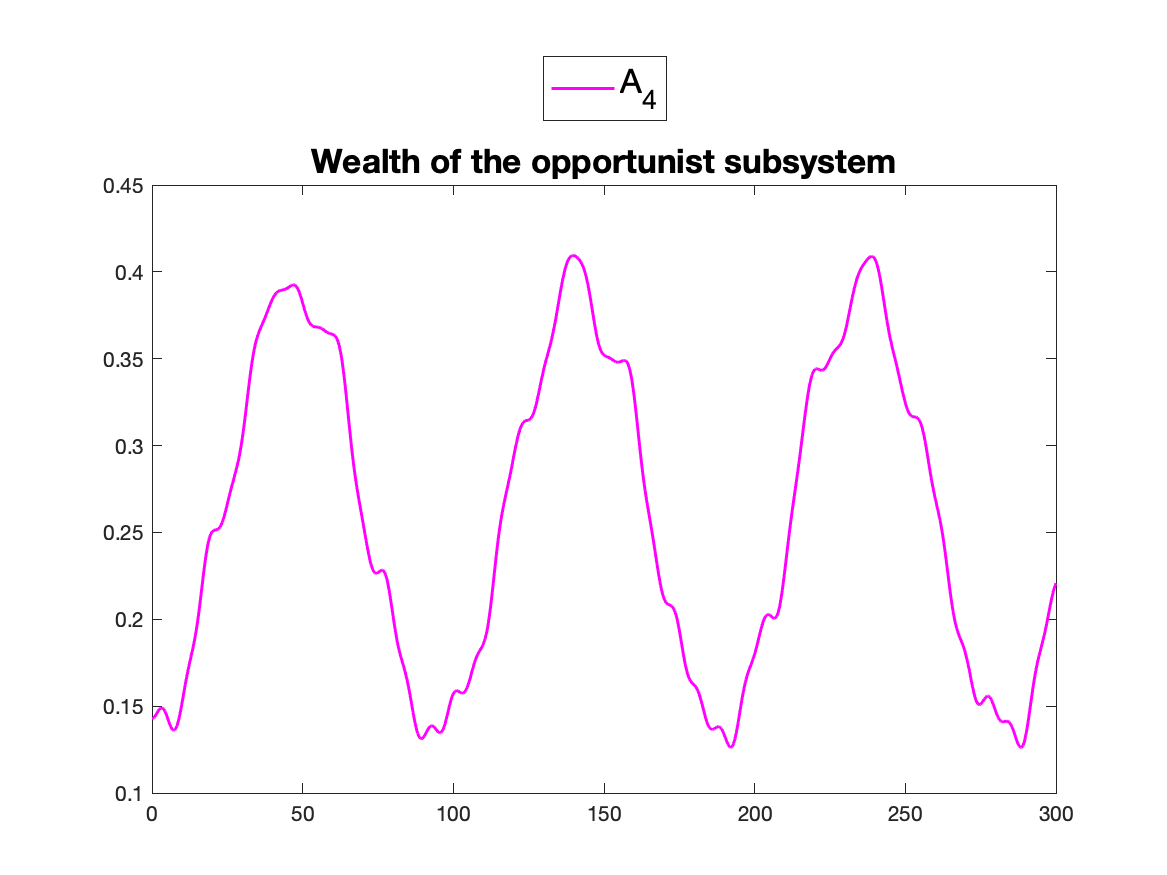}}
\subfigure[]{\includegraphics[width=0.47\textwidth]{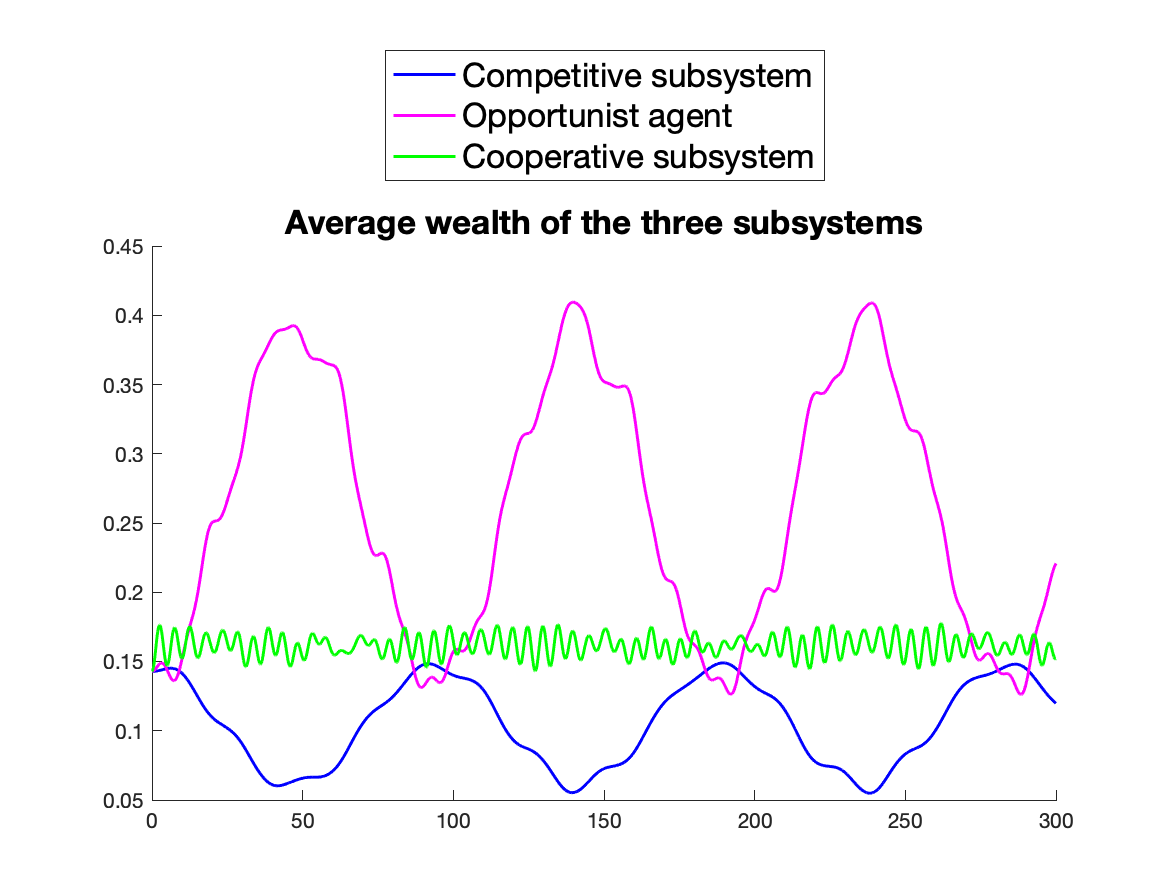}\label{fig:norule-d}}
\caption{\label{fig:norule}Time evolution of wealth of the agents as a function of time according 
to the classical Heisenberg dynamics: subfigure (a) is concerned with the competitive subsystem; subfigure (b) is concerned with the 
cooperative subsystem; subfigure (c) displays 
the wealth of the opportunist agent $A_4$; finally, subfigure (d) displays the average of the 
wealth of competitive subsystem, cooperative subsystem, and opportunist subsystem.}
\end{center}
\end{figure}

For the considered model, the evolution equations read
\[
\begin{aligned}
&\dot a_1=\textrm{i}\left(-\omega_1a_1+\lambda_{1,2}a_2+\lambda_{1,3}a_3\right),\\
&\dot a_2=\textrm{i}\left(-\omega_2a_2+\lambda_{1,2}a_1+\lambda_{2,3}a_3+\lambda_{2,4}a_4\right),\\
&\dot a_3=\textrm{i}\left(-\omega_3a_3+\lambda_{1,3}a_1+\lambda_{2,3}a_2\right),\\
&\dot a_4=\textrm{i}\left(-\omega_4a_4+\lambda_{2,4}a_2-\mu_{4,6}a_6^\dagger\right),\\
&\dot a_5=\textrm{i}\left(-\omega_5a_5-\mu_{5,6}a_6^\dagger-\mu_{5,7}a_7^\dagger\right),\\
&\dot a_6=\textrm{i}\left(-\omega_6a_6+\mu_{4,6}a_4^\dagger+\mu_{5,6}a_5^\dagger-\mu_{6,7}
a_7^\dagger\right),\\
&\dot a_7=\textrm{i}\left(-\omega_7a_7+\mu_{5,7}a_5^\dagger+\mu_{6,7}a_6^\dagger\right),
\end{aligned}
\] 
to be considered together with their adjoints.
Let us assign the following values to the parameters involved in the model: 
\[
\begin{aligned}
&\omega_1 = 0.5,\; \omega_2 = 0.45,\; \omega_3 = 0.55,\; \omega_4 = 0.3,\; 
\omega_5 = 0.65,\; \omega_6 = 0.5,\; \omega_7 = 0.7,\\
&\lambda_{1,2} = 0.1,\;  \lambda_{1,3} = 0.1,\;\lambda_{2,3} = 0.1,\;\lambda_{2,4} = 0.05,\\
&\mu_{4,6} = 0.05,\; \mu_{5,6} = 0.1,\; \mu_{5,7} = 0.1,\;\mu_{6,7} = 0.1,
\end{aligned}
\]
and let us assign at $t=0$ to every agent the same amount of wealth (equal to $1/7$).

Some comments about the choice of the values of the parameters are in order. The inertia parameters of the 
purely competitive agents are smaller than those of the purely cooperative agents, and the 
opportunist agent has the lower inertia parameter. Also, the parameter responsible for competition 
among the subsystem $\mathcal{S}_1$ is the same as the parameter responsible for cooperation among 
the subsystem $\mathcal{S}_3$; finally, the parameters entering in the competitive and cooperative 
interactions of the opportunist agent are taken smaller.
Figure~\ref{fig:norule} displays the time evolution of the wealth of the agents, that, as one expects, is never ending oscillatory.
Figure~\ref{fig:norule-d} displays the average values of wealth of the three subsystems as a function of time: the amplitude of the oscillations is quite small for the cooperative subsystem, whereas the opportunist agent experiences the maximum amplitude of oscillations. Moreover, the maximum of wealth of the opportunist agent is higher than those of
cooperative and competitive subsystems; finally, the minimum wealth of opportunist agent is close to the maximum wealth of the competitive subsystem and to the average wealth of cooperative system.

Situation drastically changes if we apply periodically a rule, that is, if the agents are able to change their attitudes according to the evolution of their wealth, as shown in the next Subsection.

\subsection{Numerical simulations with the $(\mathcal{H},\rho)$--induced dynamics approach}
The description of the dynamics can be enriched by introducing a 
\emph{rule} able to include in the model some effects that can not be easily implemented in the definition 
of $\mathcal{H}$, unless we do not assume a time--dependent Hamiltonian or the existence of a reservoir.  The rule we use is detailed below. 

\begin{figure}
\begin{center}
\subfigure[]{\includegraphics[width=0.47\textwidth]{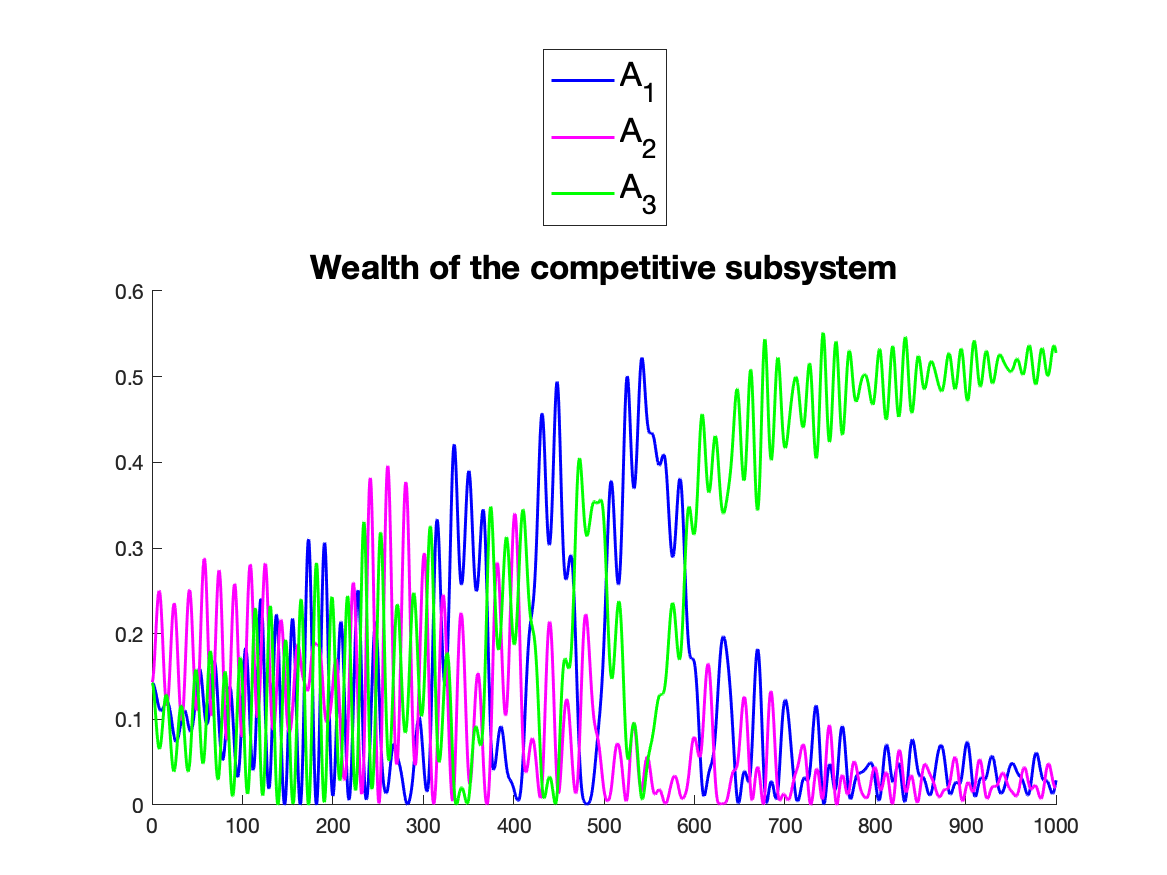}}
\subfigure[]{\includegraphics[width=0.47\textwidth]{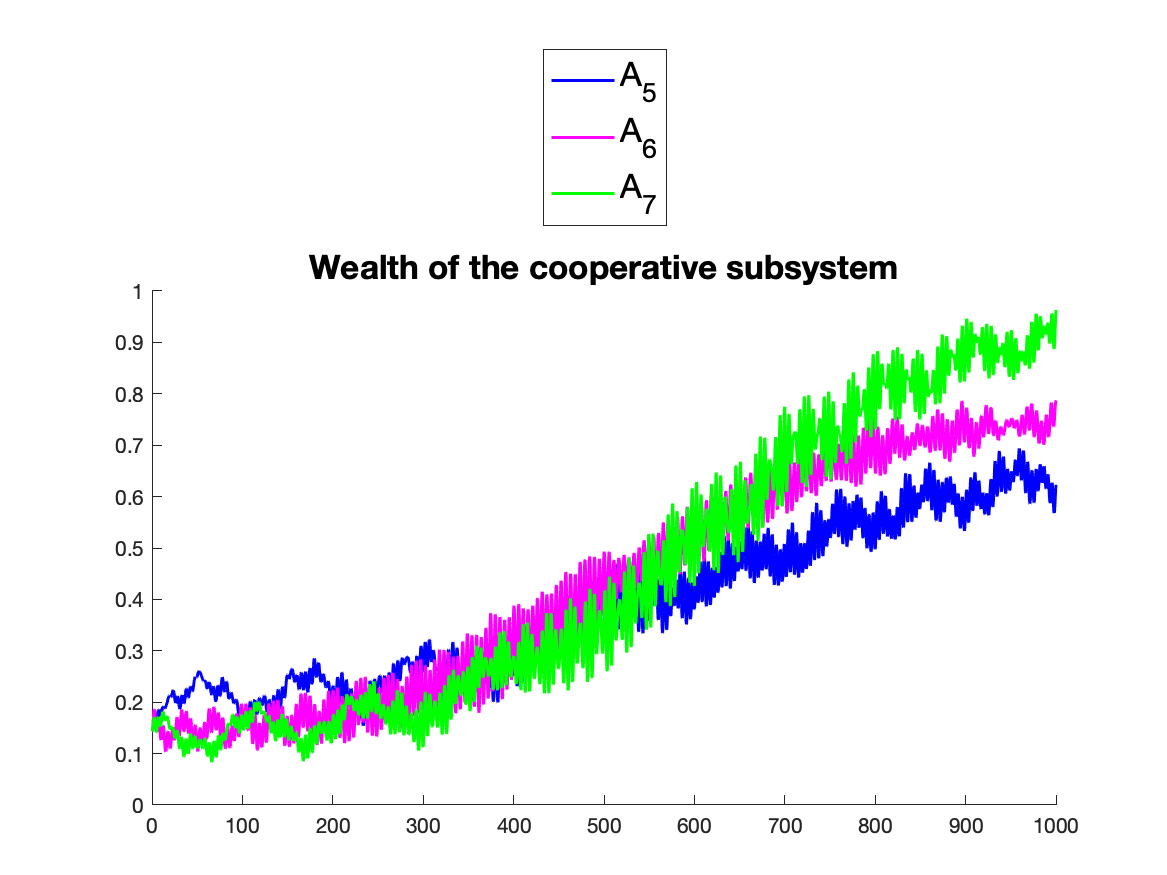}}\\
\subfigure[]{\includegraphics[width=0.47\textwidth]{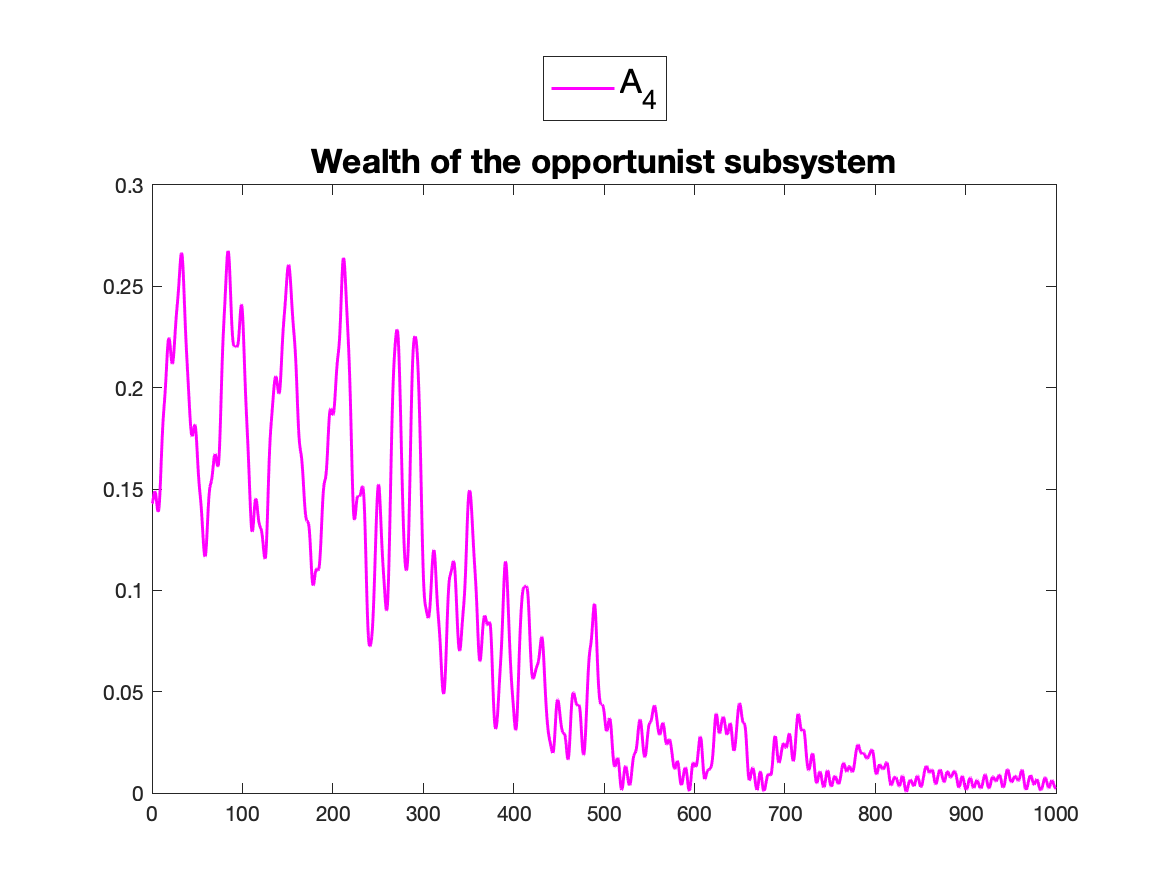}}
\subfigure[]{\includegraphics[width=0.47\textwidth]{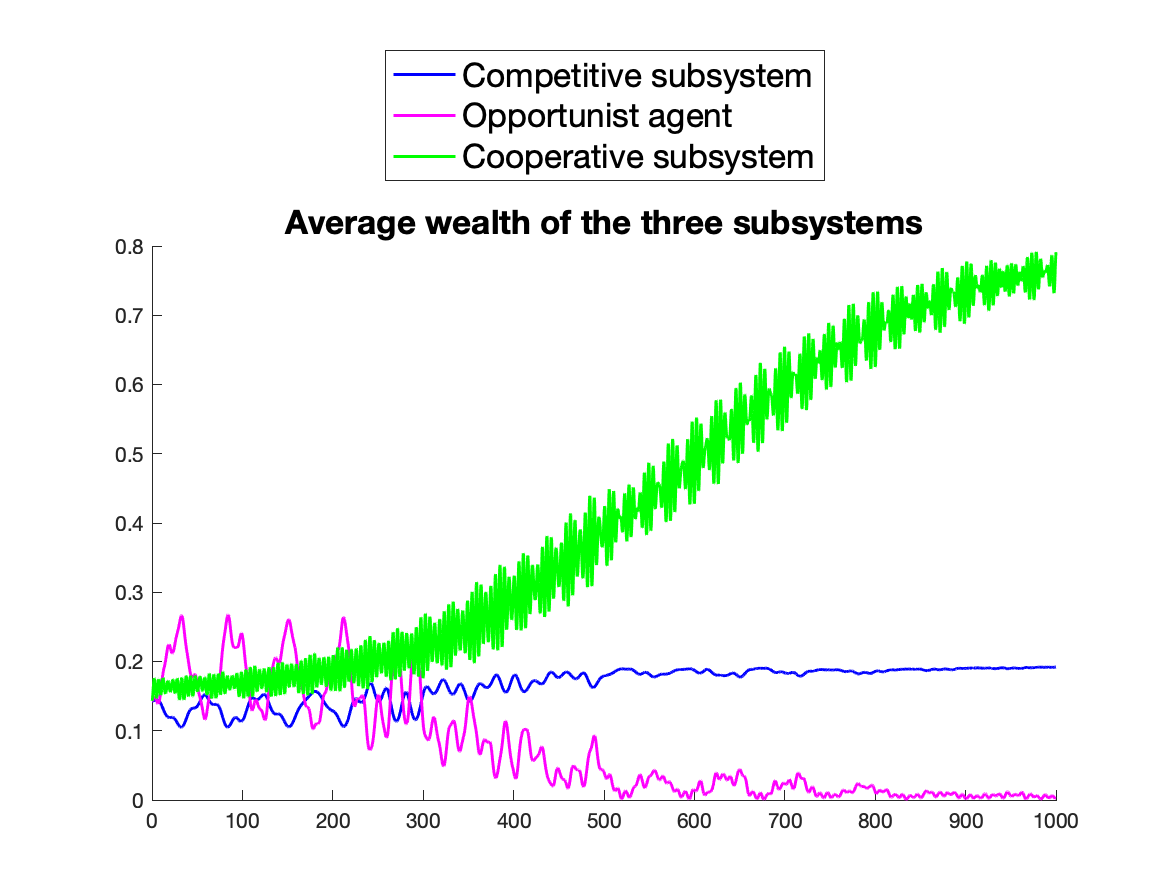}}
\caption{\label{fig:rulet1}Time evolution of wealth of the agents as a function of time using the 
$(\mathcal{H},\rho)$--induced dynamics approach with $\tau=1$: subfigure (a) is concerned with the 
competitive subsystem, subfigure (b) is 
concerned with the cooperative subsystem, 
subfigure (c) displays the wealth of the opportunist agent; finally, subfigure (d) displays 
the average of the wealth of competitive subsystem, cooperative subsystem, and 
opportunist subsystem.}
\end{center}
\end{figure}

\begin{figure}
\begin{center}
\subfigure[]{\includegraphics[width=0.47\textwidth]{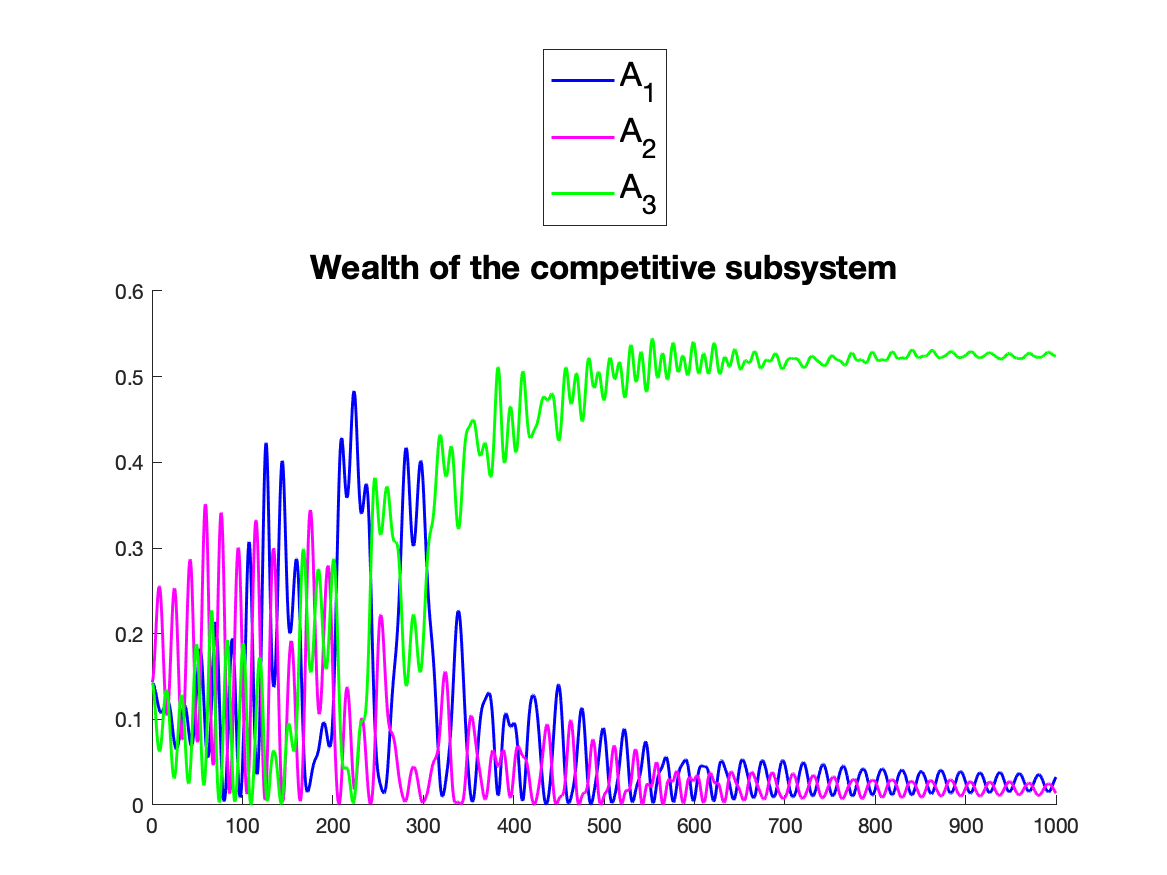}}
\subfigure[]{\includegraphics[width=0.47\textwidth]{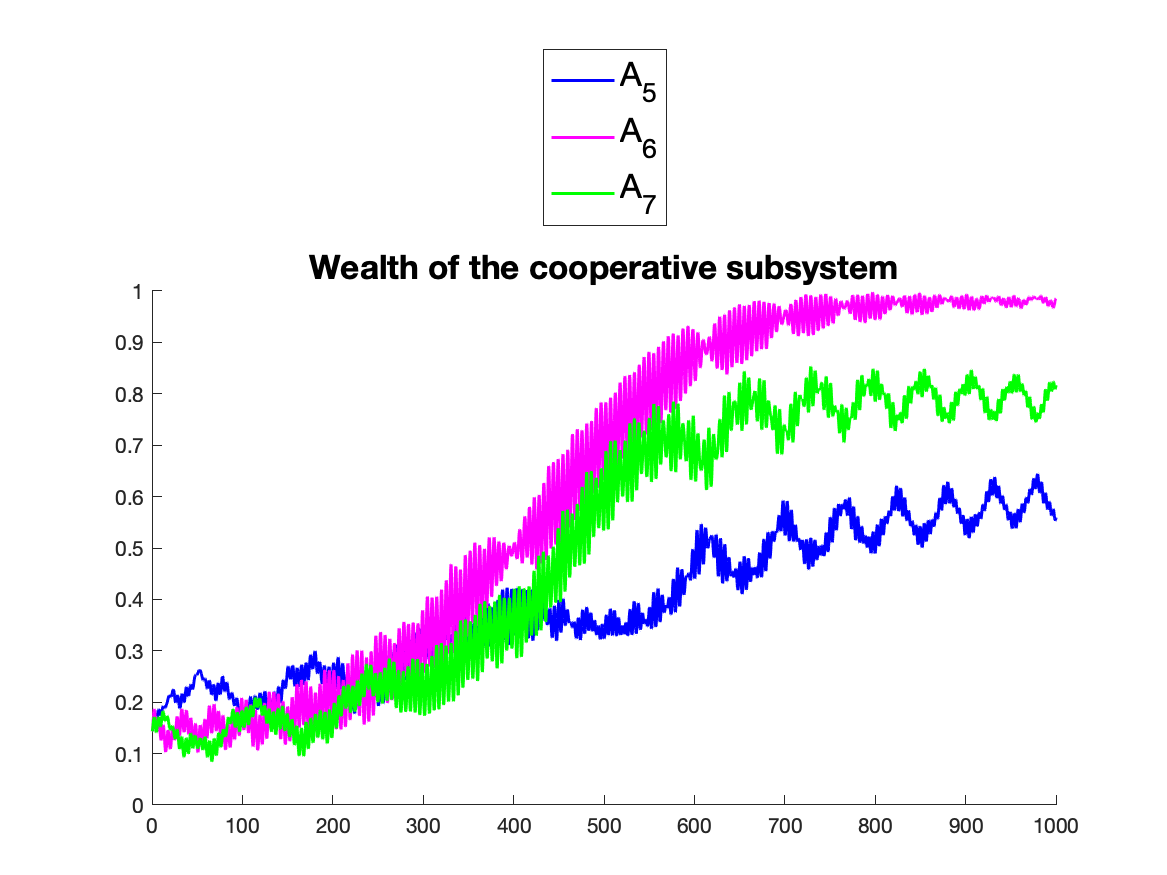}}\\
\subfigure[]{\includegraphics[width=0.47\textwidth]{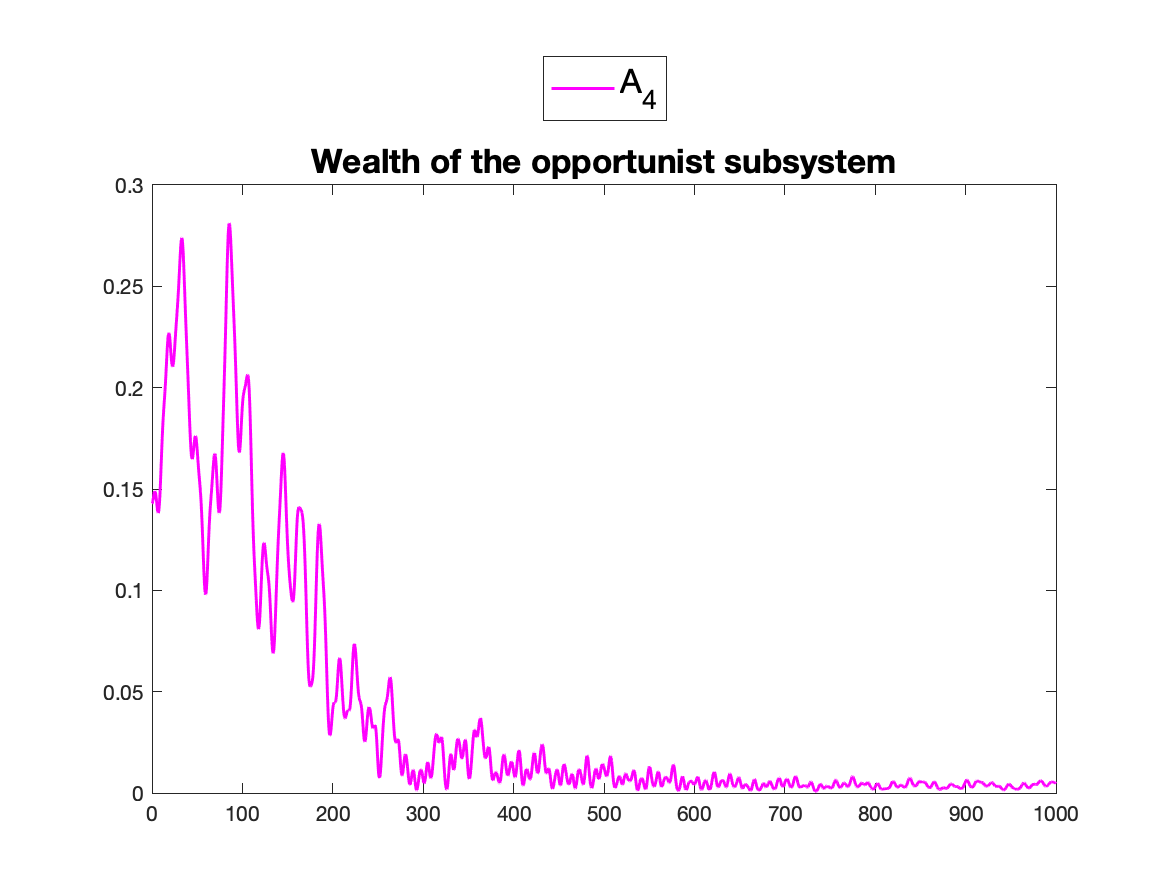}}
\subfigure[]{\includegraphics[width=0.47\textwidth]{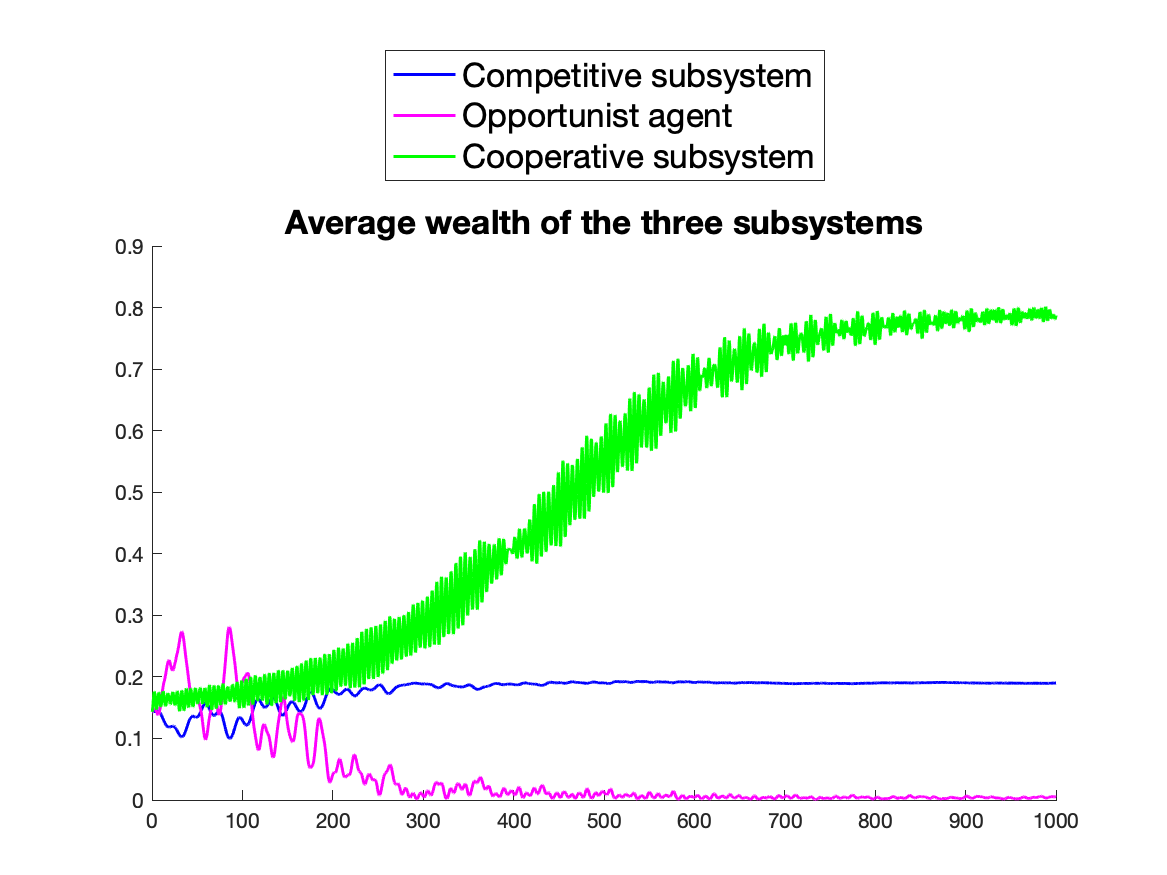}}
\caption{\label{fig:rulet2}Time evolution of wealth of the agents as a function of time using the 
$(\mathcal{H},\rho)$--induced dynamics approach with $\tau=2$: subfigure (a) is concerned with the 
competitive subsystem, subfigure (b) is 
concerned with the cooperative subsystem, 
subfigure (c) displays the wealth of the opportunist agent; finally, subfigure (d) displays 
the average of the wealth of competitive subsystem, cooperative subsystem, and 
opportunist subsystem.}
\end{center}
\end{figure}

\begin{figure}
\begin{center}
\subfigure[]{\includegraphics[width=0.47\textwidth]{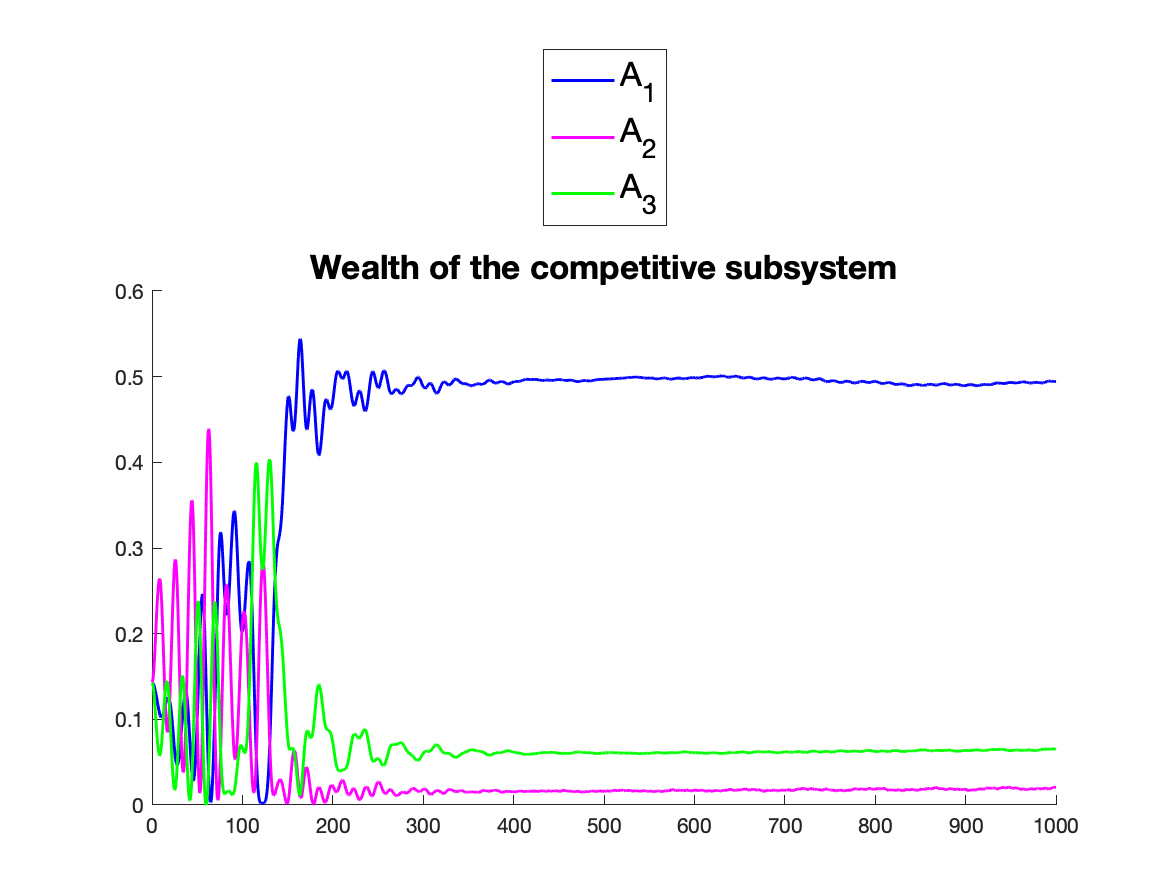}}
\subfigure[]{\includegraphics[width=0.47\textwidth]{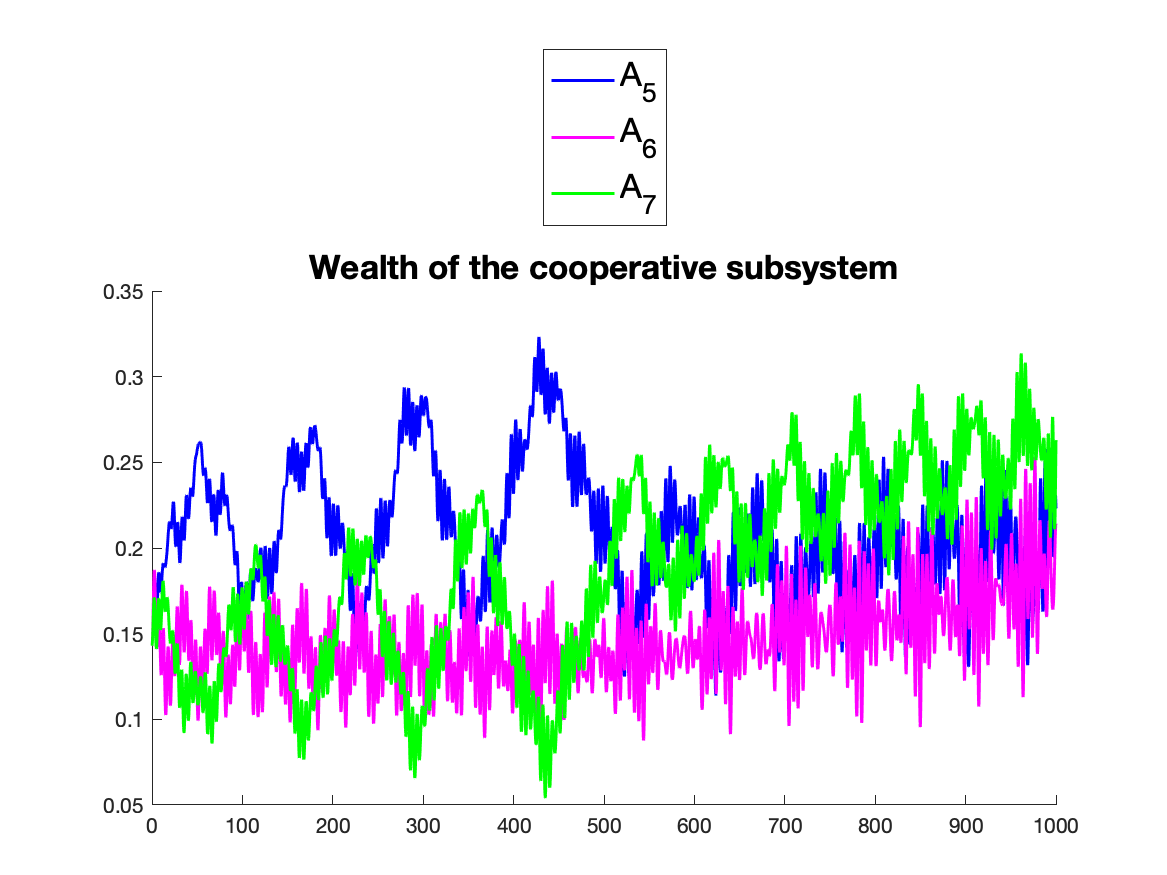}}\\
\subfigure[]{\includegraphics[width=0.47\textwidth]{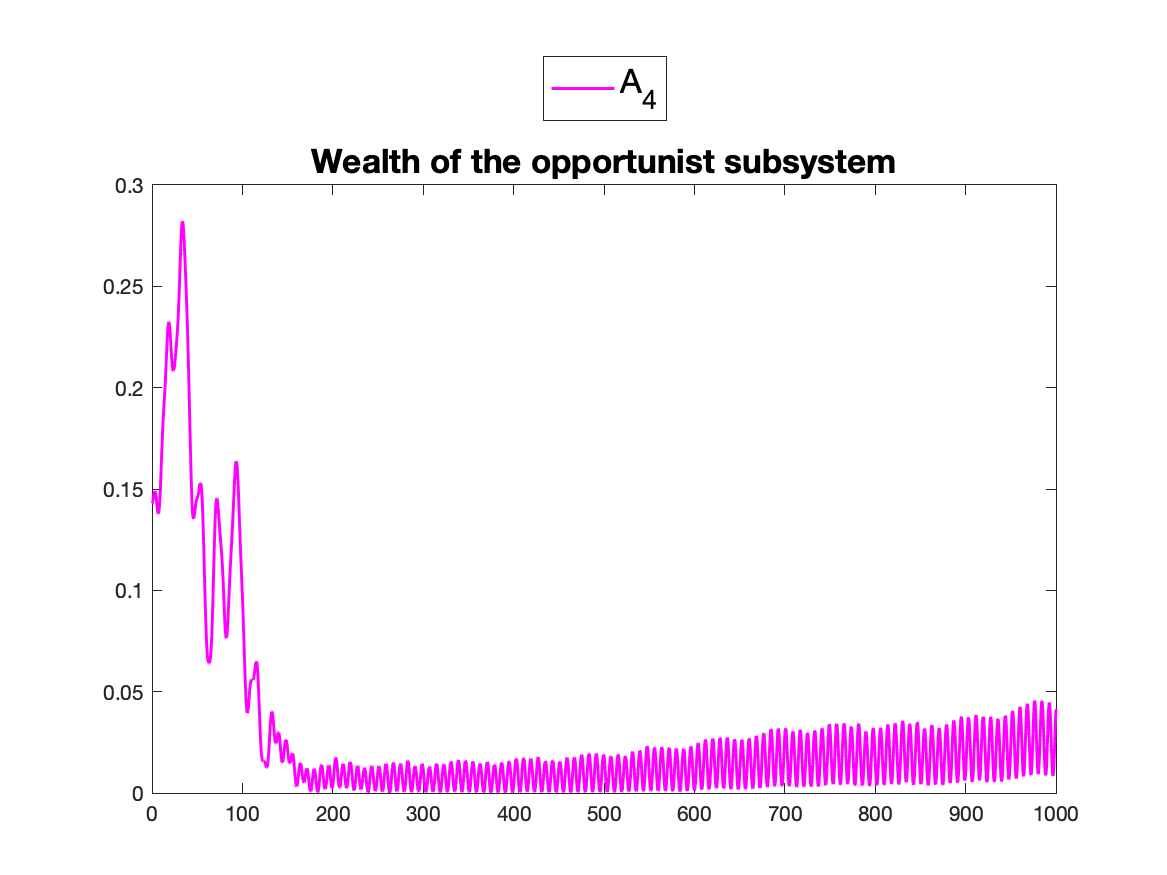}}
\subfigure[]{\includegraphics[width=0.47\textwidth]{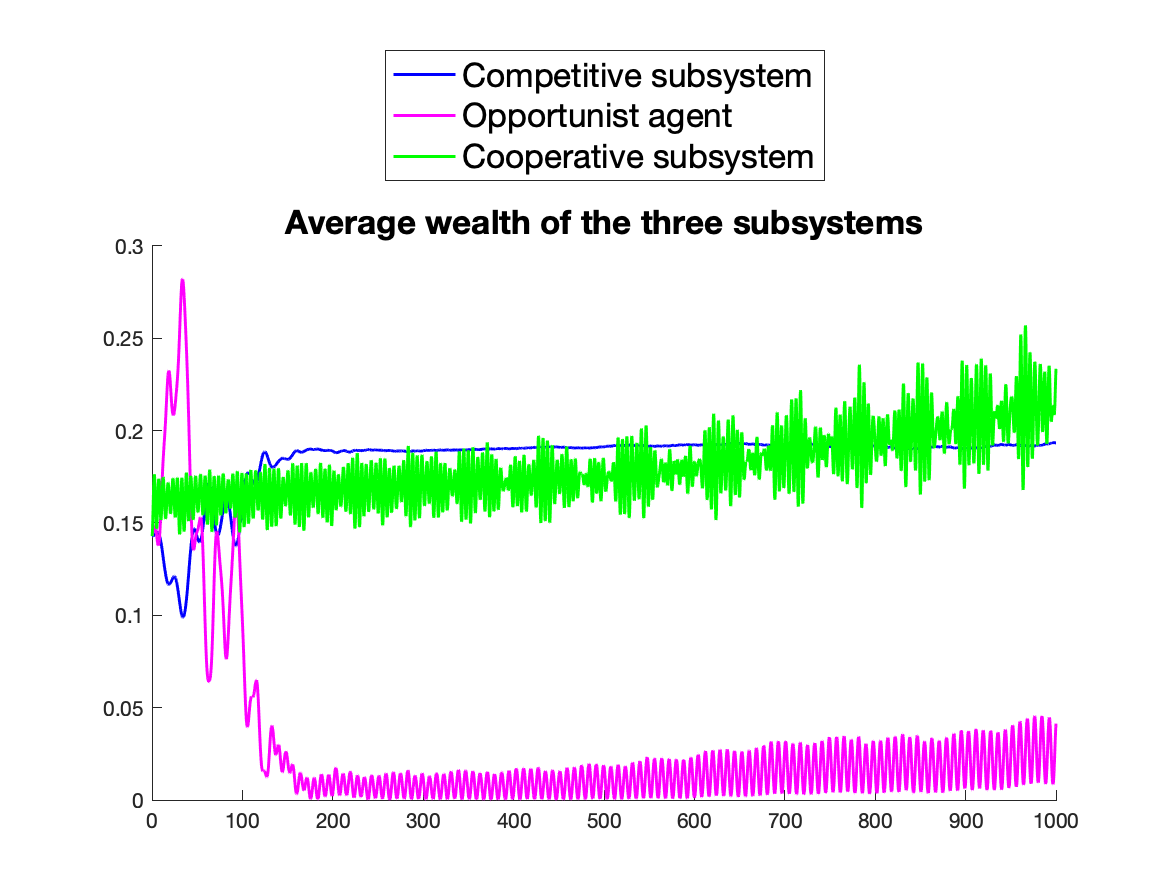}}
\caption{\label{fig:rulet4}Time evolution of wealth of the agents as a function of time using the 
$(\mathcal{H},\rho)$--induced dynamics approach with $\tau=4$: subfigure (a) is concerned with the 
competitive subsystem, subfigure (b) is 
concerned with the cooperative subsystem, 
subfigure (c) displays the wealth of the opportunist agent; finally, subfigure (d) displays 
the average of the wealth of competitive subsystem, cooperative subsystem, and 
opportunist subsystem.}
\end{center}
\end{figure}

\begin{figure}
\begin{center}
\subfigure[]{\includegraphics[width=0.47\textwidth]{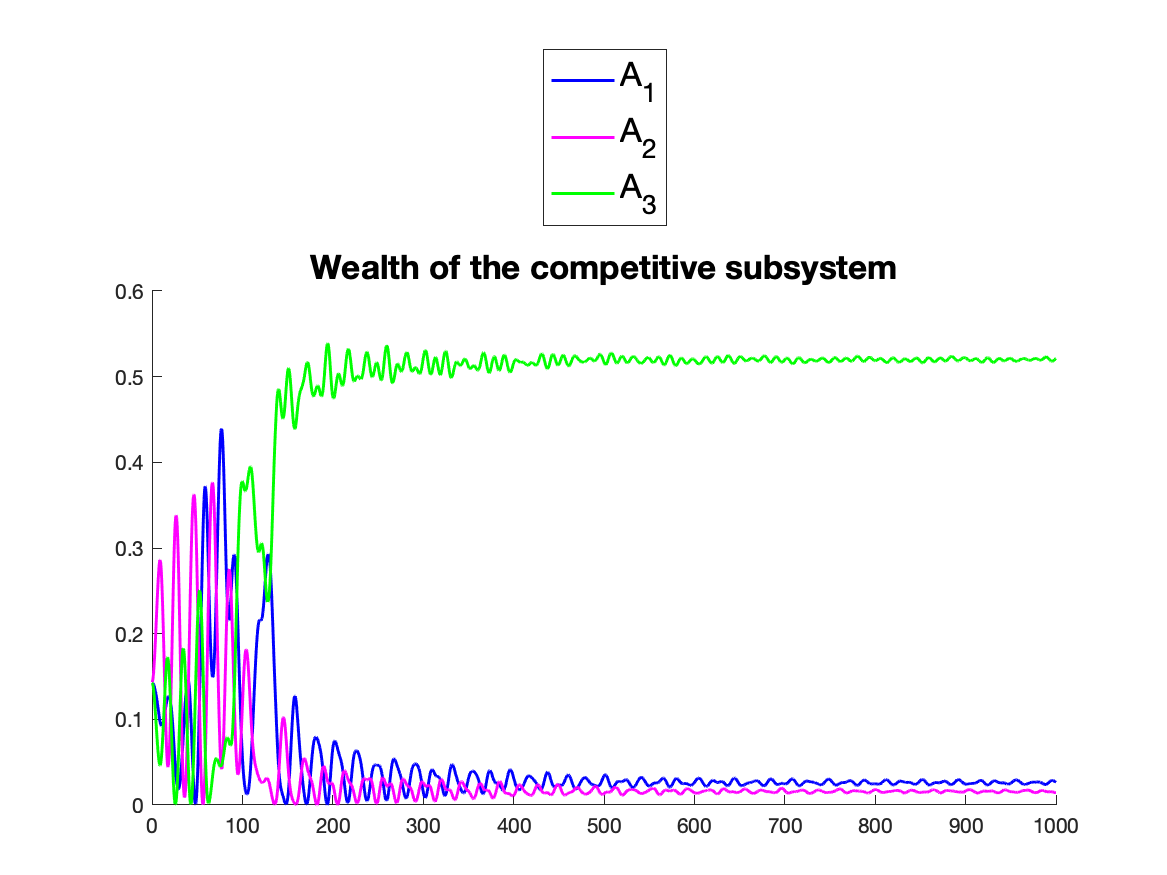}}
\subfigure[]{\includegraphics[width=0.47\textwidth]{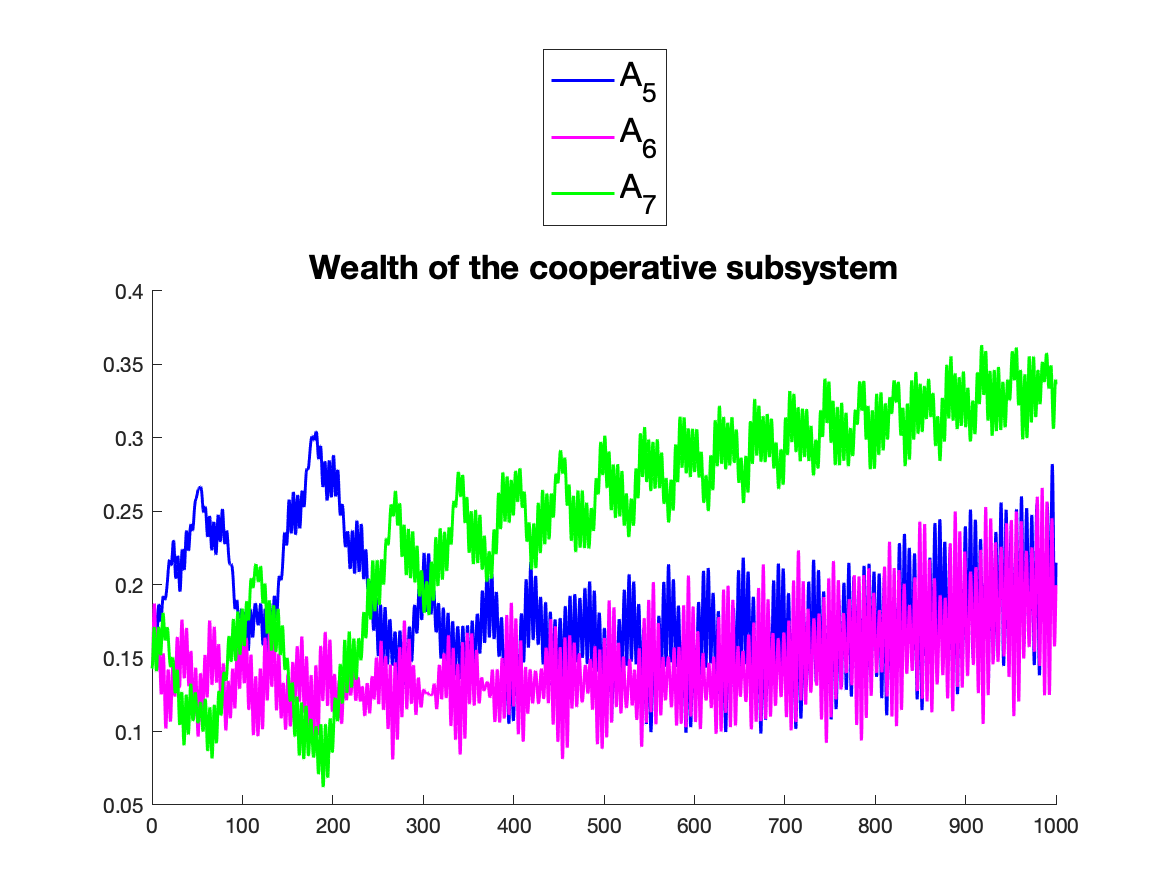}}\\
\subfigure[]{\includegraphics[width=0.47\textwidth]{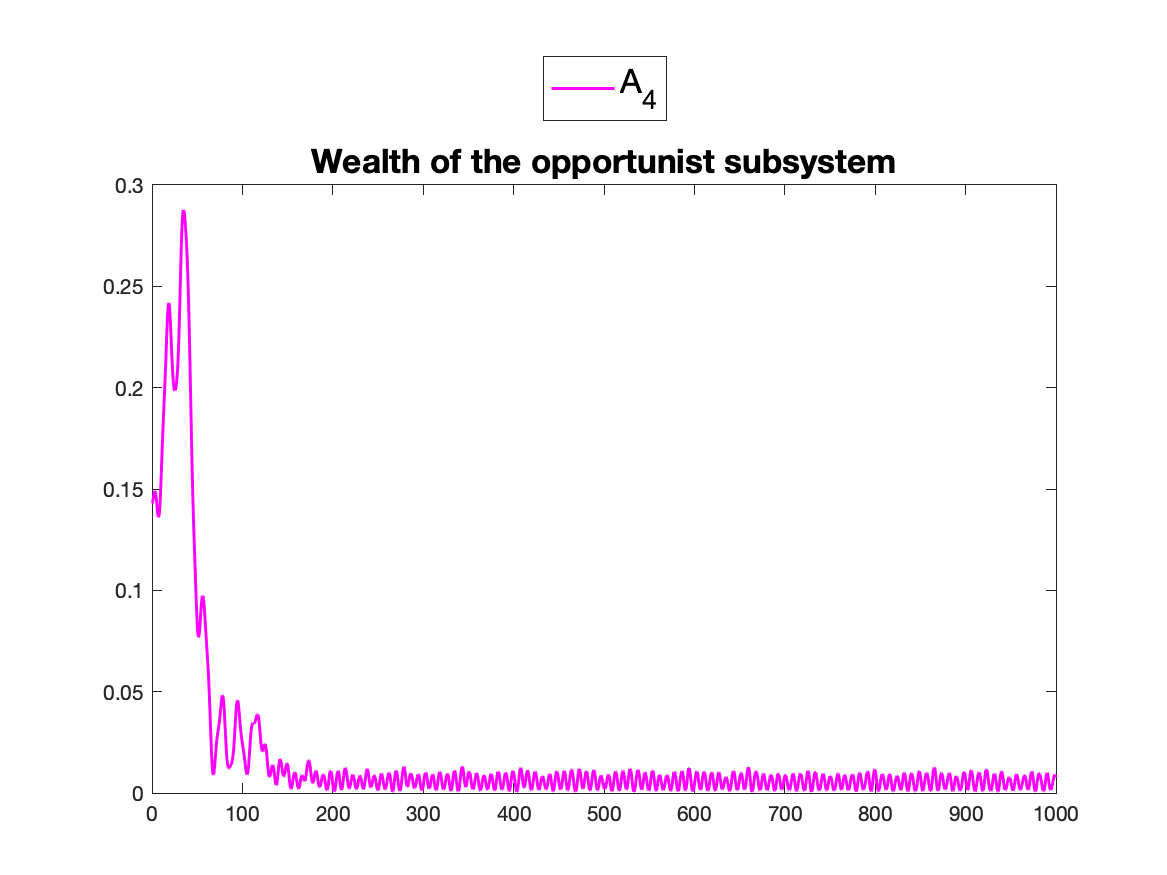}}
\subfigure[]{\includegraphics[width=0.47\textwidth]{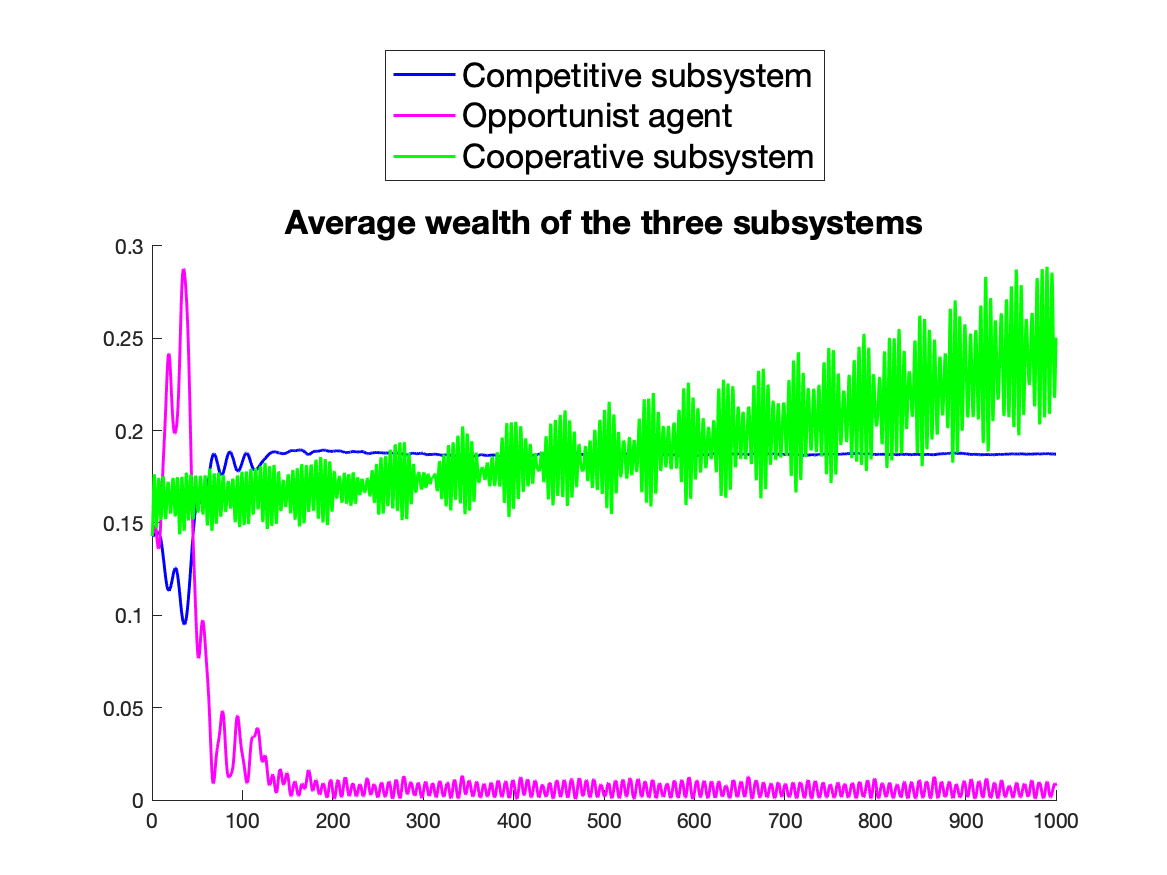}}
\caption{\label{fig:rulet8}Time evolution of wealth of the agents as a function of time using the 
$(\mathcal{H},\rho)$--induced dynamics approach with $\tau=8$: subfigure (a) is concerned with the 
competitive subsystem, subfigure (b) is 
concerned with the cooperative subsystem, 
subfigure (c) displays the wealth of the opportunist agent; finally, subfigure (d) displays 
the average of the wealth of competitive subsystem, cooperative subsystem, and 
opportunist subsystem.}
\end{center}
\end{figure}

Fixing a value for $\tau$ (the choice of $\tau$ plays a role in the dynamics), let us define
\[
\delta_{j}^{(k)}=n_{j}(k\tau)-n_{j}((k-1)\tau), \qquad j=1,\ldots,7.
\]
At the instants $k\tau$ ($k=1,2,\ldots)$ we modify the inertia parameters as follows:
\[
\omega_{j}= \omega_{j}(1+\delta_j^{(k)}), \qquad j=1,\ldots,7;
\]
therefore, the inertia parameter of the agent $A_j$ increases (decreases) if its wealth state 
in the subinterval of length $\tau$ increases (decreases); due to the meaning of the inertia 
parameters, this means that an agent increasing  its wealth lowers its tendency to change. 
On the contrary, an agent undergoing to a decrease of its wealth is induced to become less 
conservative.

Figures~\ref{fig:rulet1}, \ref{fig:rulet2}, \ref{fig:rulet4} and \ref{fig:rulet8} show the results when the 
$(\mathcal{H},\rho)$--induced dynamics approach is used with $\tau=1$, $\tau=2$, $\tau=4$ and $\tau=8$, respectively. 
We observe that the transient behavior of the time evolution changes with $\tau$ and there is a general damping of the amplitudes of the oscillations. Moreover, if we look at the subfigures (d), we observe  that, for the values of $\tau$ here considered, as $t$ increases, the cooperative subsystem gains the higher amount of wealth, whereas the wealth of the competitive subsystem is almost conserved (at least for $\tau=1$ and $\tau=2$), and the wealth of the opportunist subsystem experiences the greatest loss. For higher values of $\tau$ ($=4, 8$), the average wealth of competitive subsystem 
is higher than that of the opportunist agent.

\section{An extended spatial model}
\label{sec:manymodemodel}
Let us consider a more sophisticated system made by $N$ agents; each agent is located in a cell of  
a one--dimensional torus partitioned in $N$ cells, so that the cell 1 is adjacent to the cell $N$. 
Moreover, the distance between adjacent cells is assumed to be 1, whereupon the maximum distance 
between the cells is $d_{\hbox{max}}=\left\lfloor N/2\right\rfloor$.

\begin{figure}
\begin{center}
\subfigure[]{\includegraphics[width=0.47\textwidth]{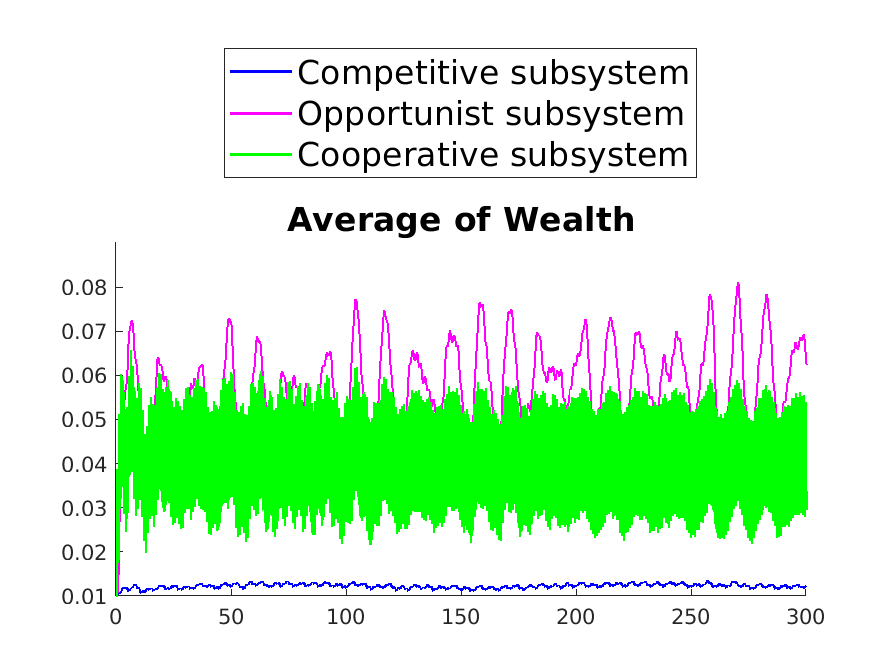}}
\subfigure[]{\includegraphics[width=0.47\textwidth]{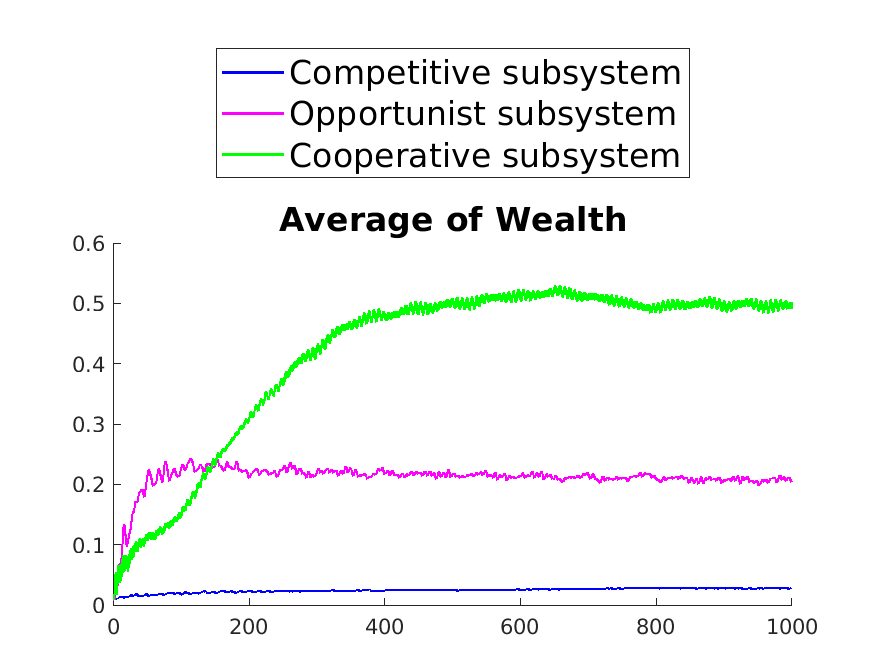}}\\
\subfigure[]{\includegraphics[width=0.47\textwidth]{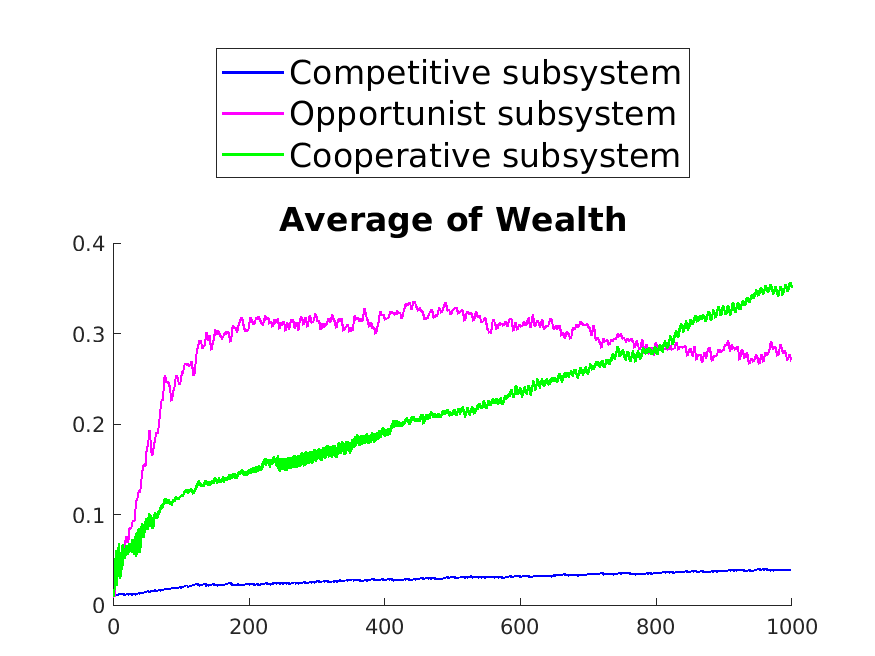}}
\subfigure[]{\includegraphics[width=0.47\textwidth]{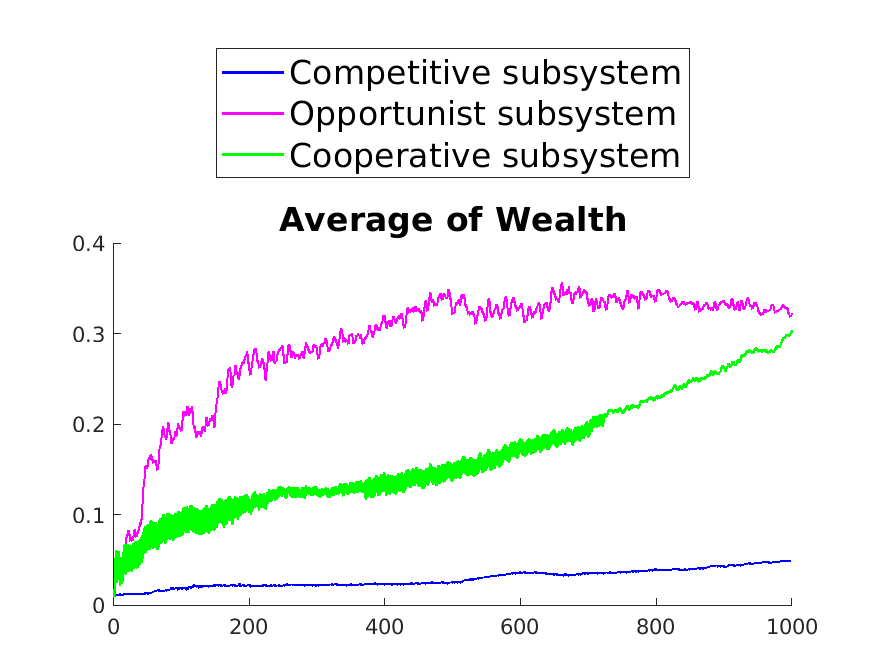}}
\caption{\label{fig:wealth100-10}Time evolution of wealth of the system with 100 agents: the 
cooperative and competitive subgroups contain 45 agents and the opportunist 
subgroup 10 agents; subfigure (a) displays the results without using the rule, subfigures (b), (c) 
and (d) the results using the $(\mathcal{H},\rho)$--induced dynamics approach with $\tau=1$, $
\tau=2$ and $\tau=4$, respectively.}
\end{center}
\end{figure}

\begin{figure}
\begin{center}
\subfigure[]{\includegraphics[width=0.47\textwidth]{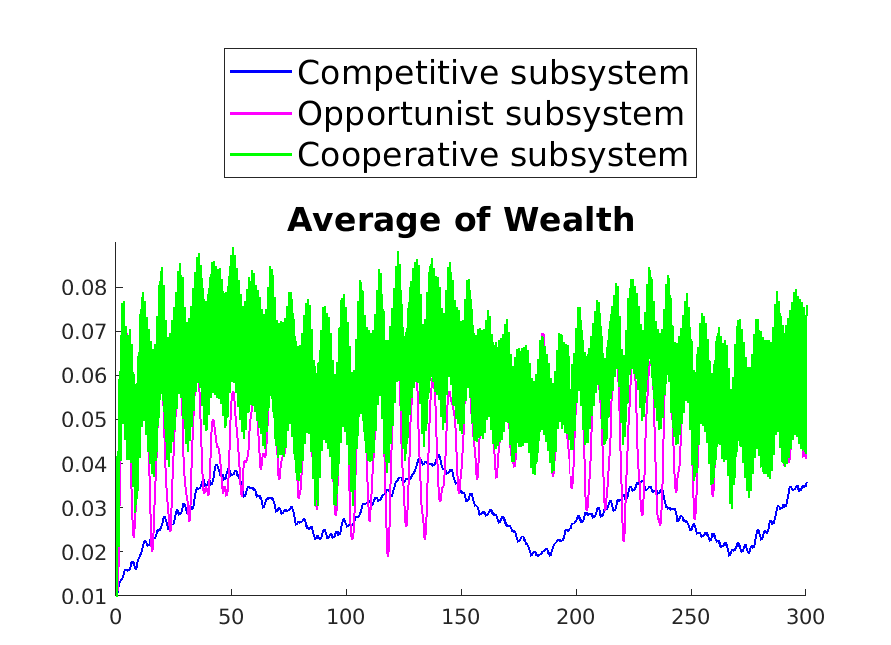}}
\subfigure[]{\includegraphics[width=0.47\textwidth]{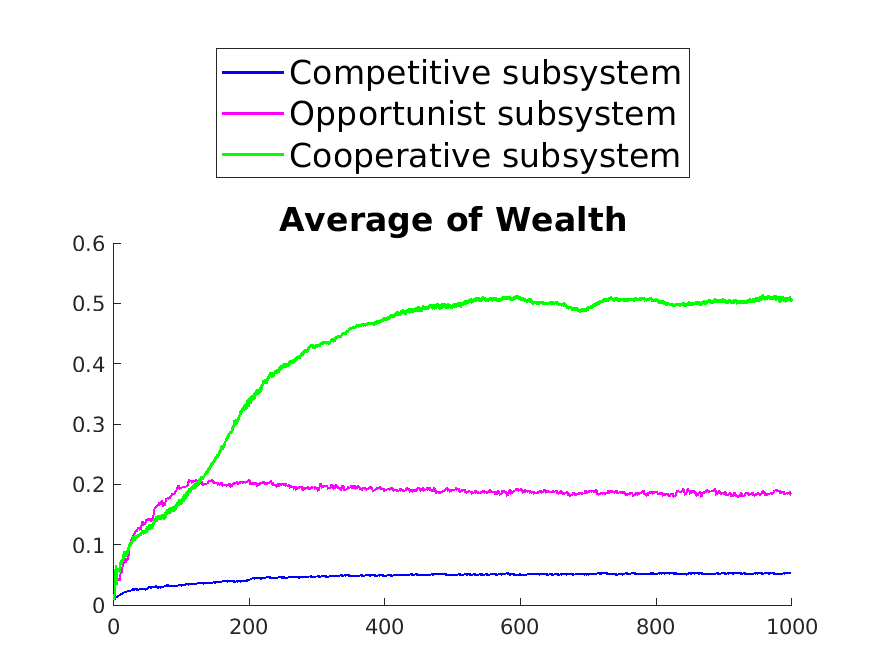}}\\
\subfigure[]{\includegraphics[width=0.47\textwidth]{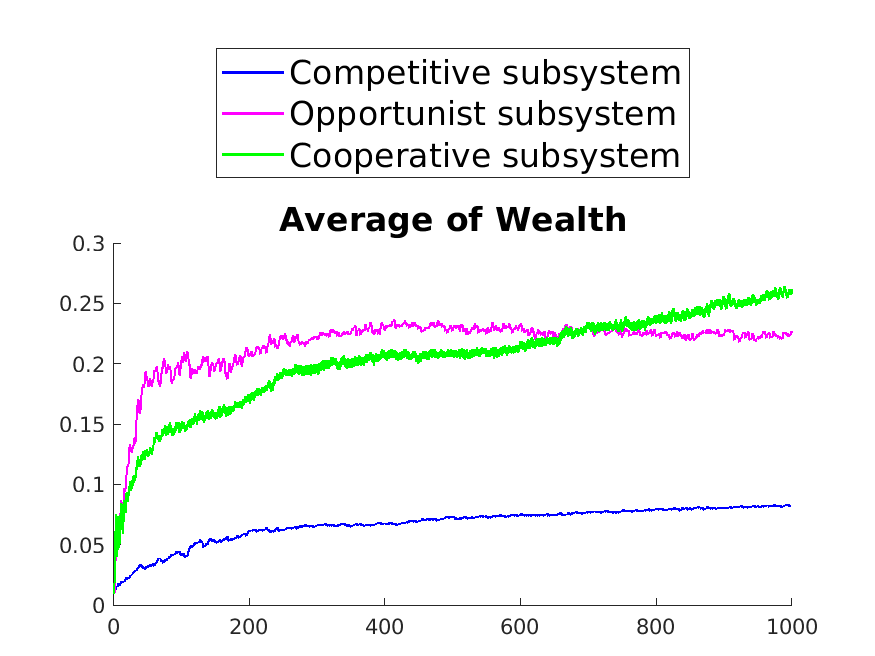}}
\subfigure[]{\includegraphics[width=0.47\textwidth]{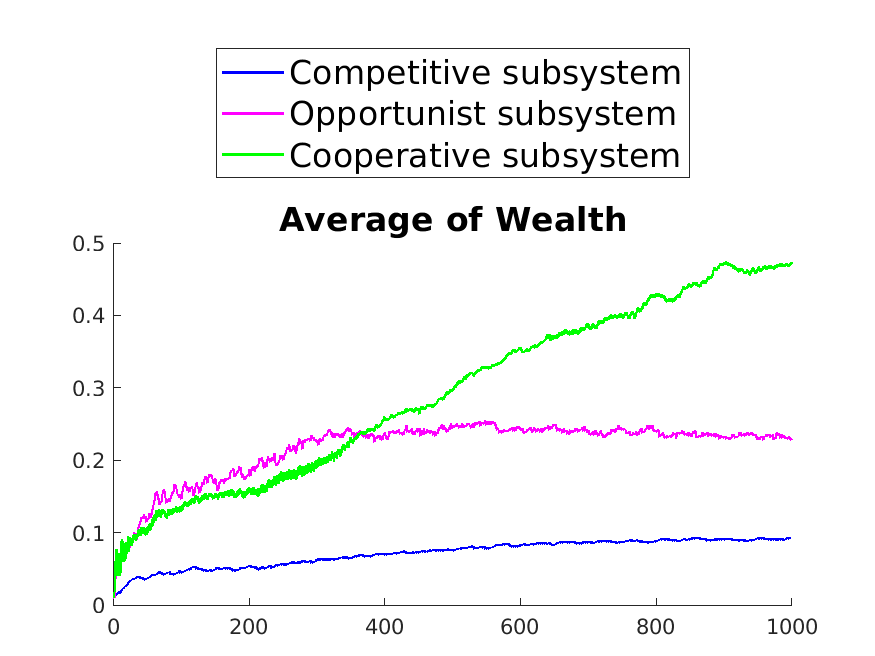}}
\caption{\label{fig:wealth100-16}Time evolution of wealth of the system with 100 agents: the 
cooperative and competitive subgroups contain 42 agents and the opportunist 
subgroup 16 agents; subfigure (a) displays the results without using the rule, subfigures (b), (c) 
and (d) the results using the $(\mathcal{H},\rho)$--induced dynamics approach with $\tau=1$, $
\tau=2$ and $\tau=4$, respectively.}
\end{center}
\end{figure}

\begin{figure}
\begin{center}
\subfigure[]{\includegraphics[width=0.47\textwidth]{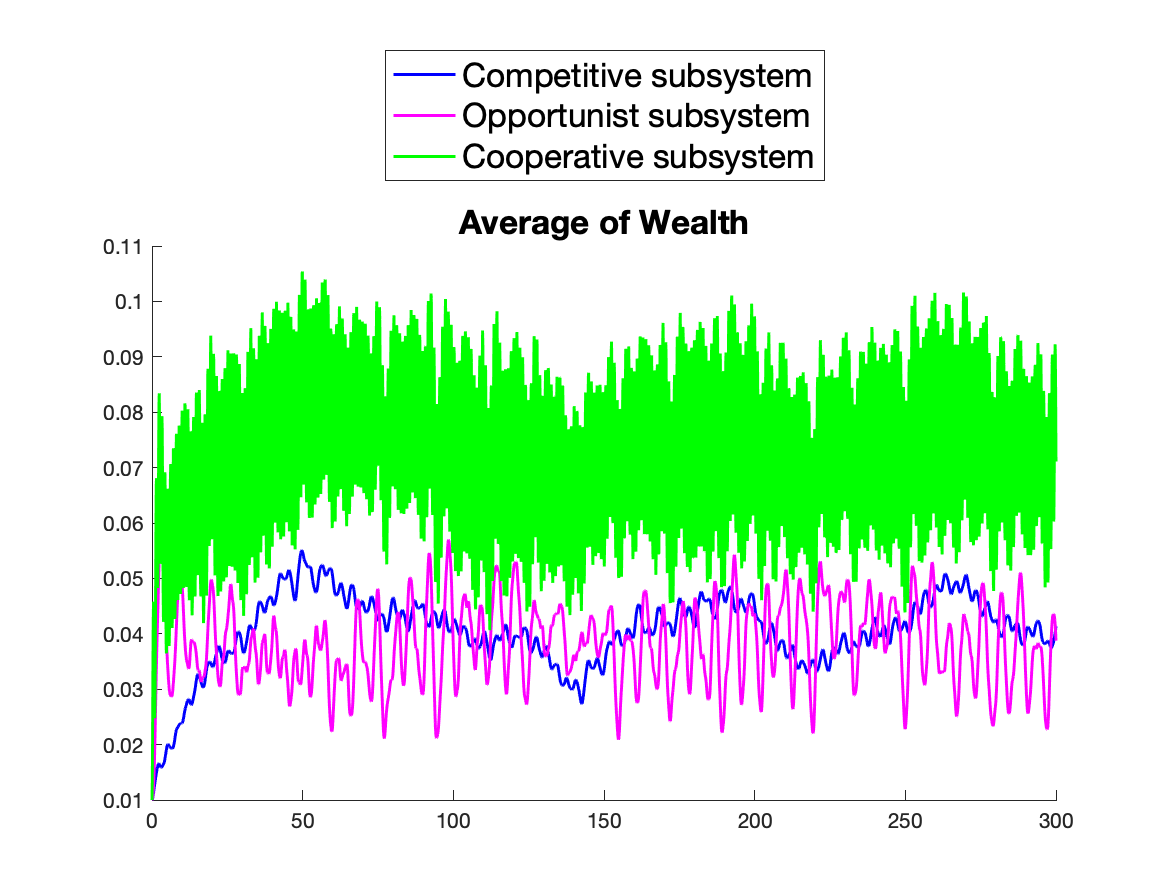}}
\subfigure[]{\includegraphics[width=0.47\textwidth]{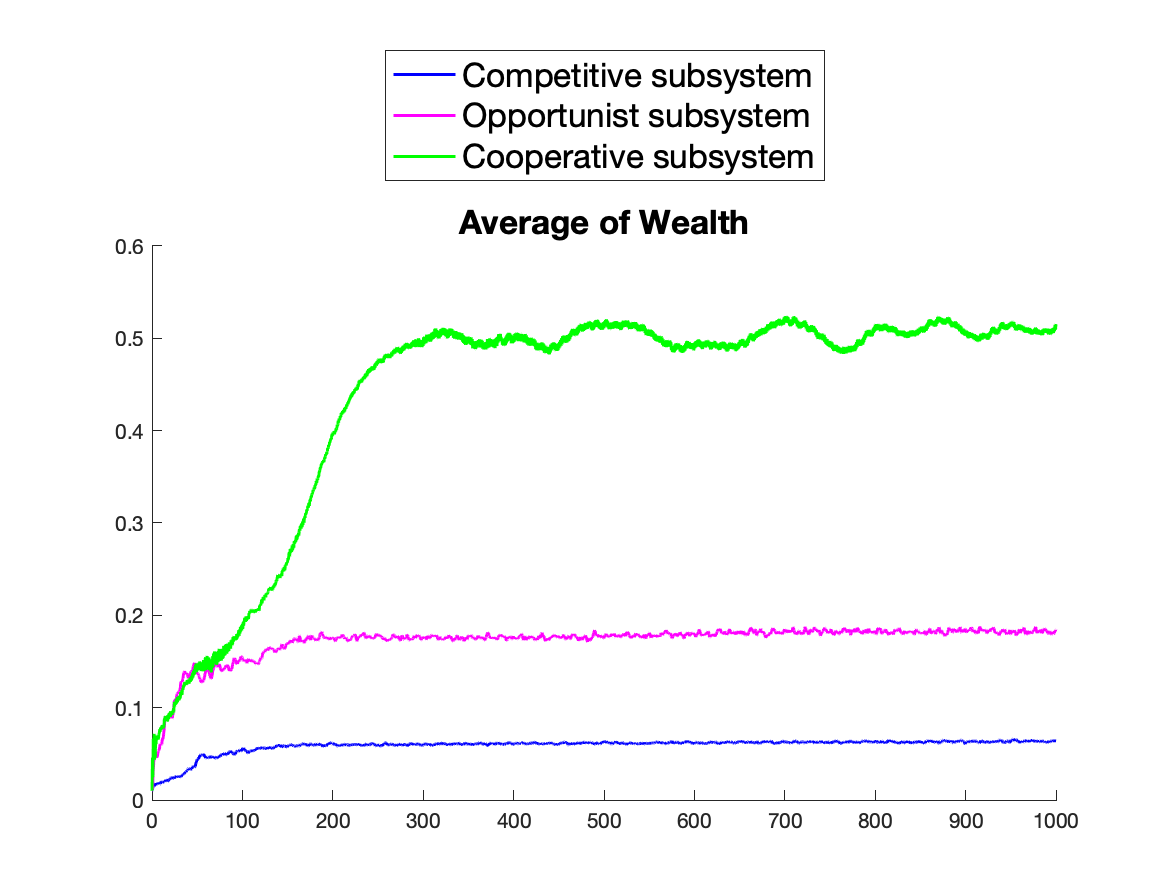}}\\
\subfigure[]{\includegraphics[width=0.47\textwidth]{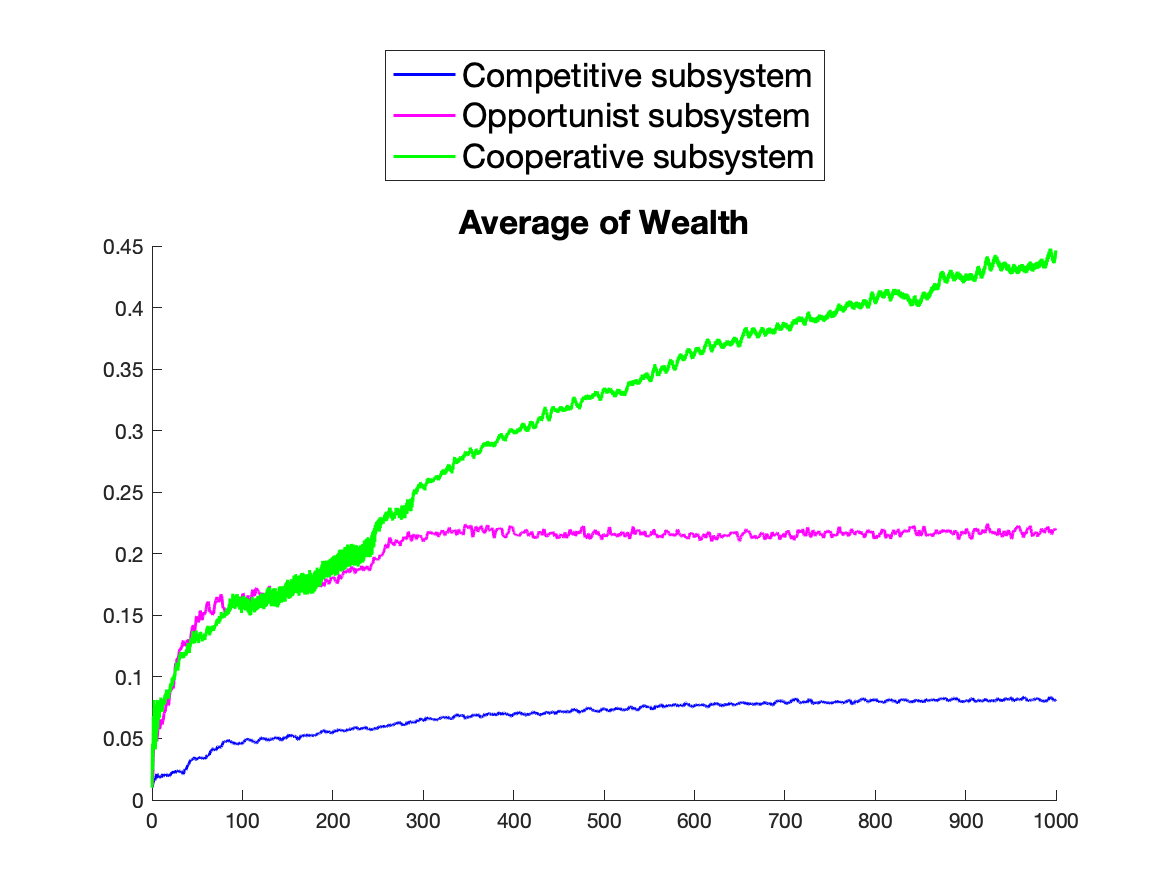}}
\subfigure[]{\includegraphics[width=0.47\textwidth]{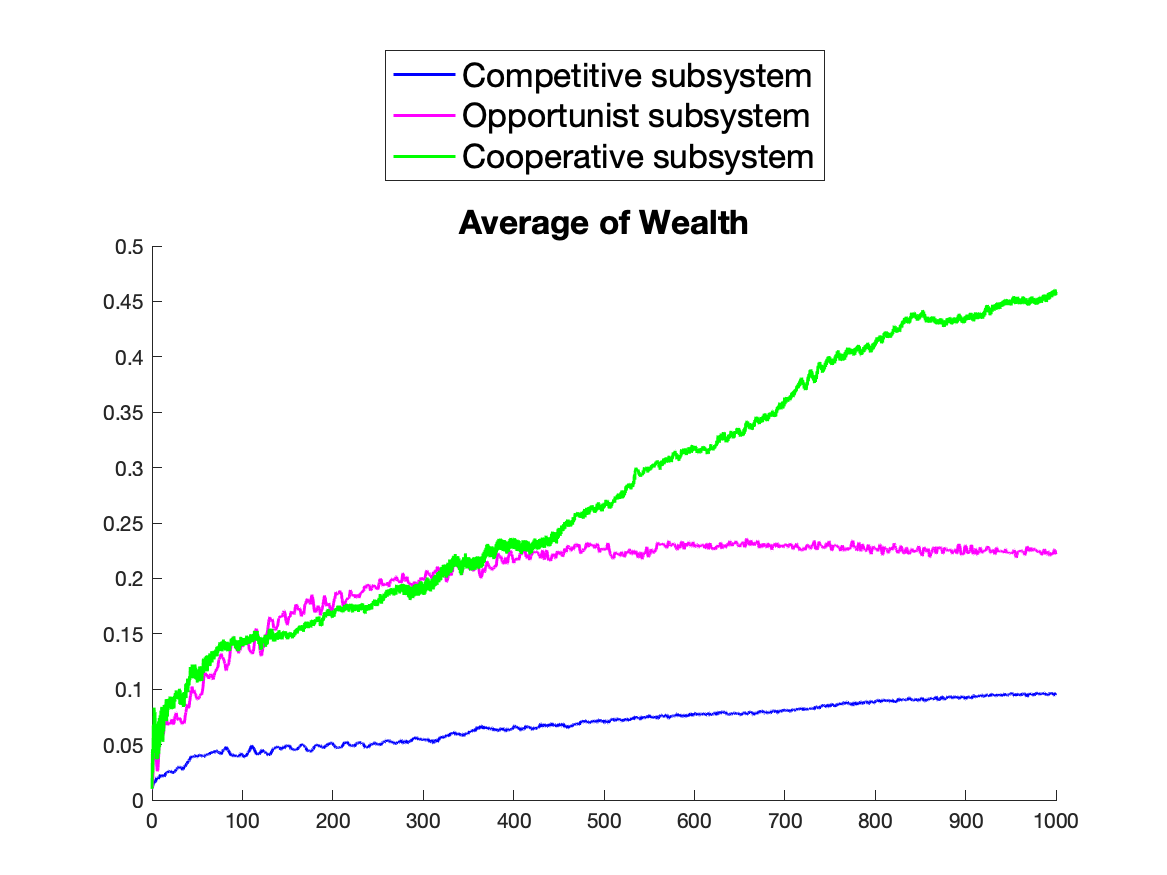}}
\caption{\label{fig:wealth100-20}Time evolution of wealth of the system with 100 agents: the 
cooperative and competitive subgroups contain 40 agents and the opportunist 
subgroup 20 agents; subfigure (a) displays the results without using the rule, subfigures (b), (c) 
and (d) the results using the $(\mathcal{H},\rho)$--induced dynamics approach with $\tau=1$, 
$\tau=2$ and $\tau=4$, respectively.}
\end{center}
\end{figure}

\begin{figure}
\begin{center}
\subfigure[]{\includegraphics[width=0.47\textwidth]{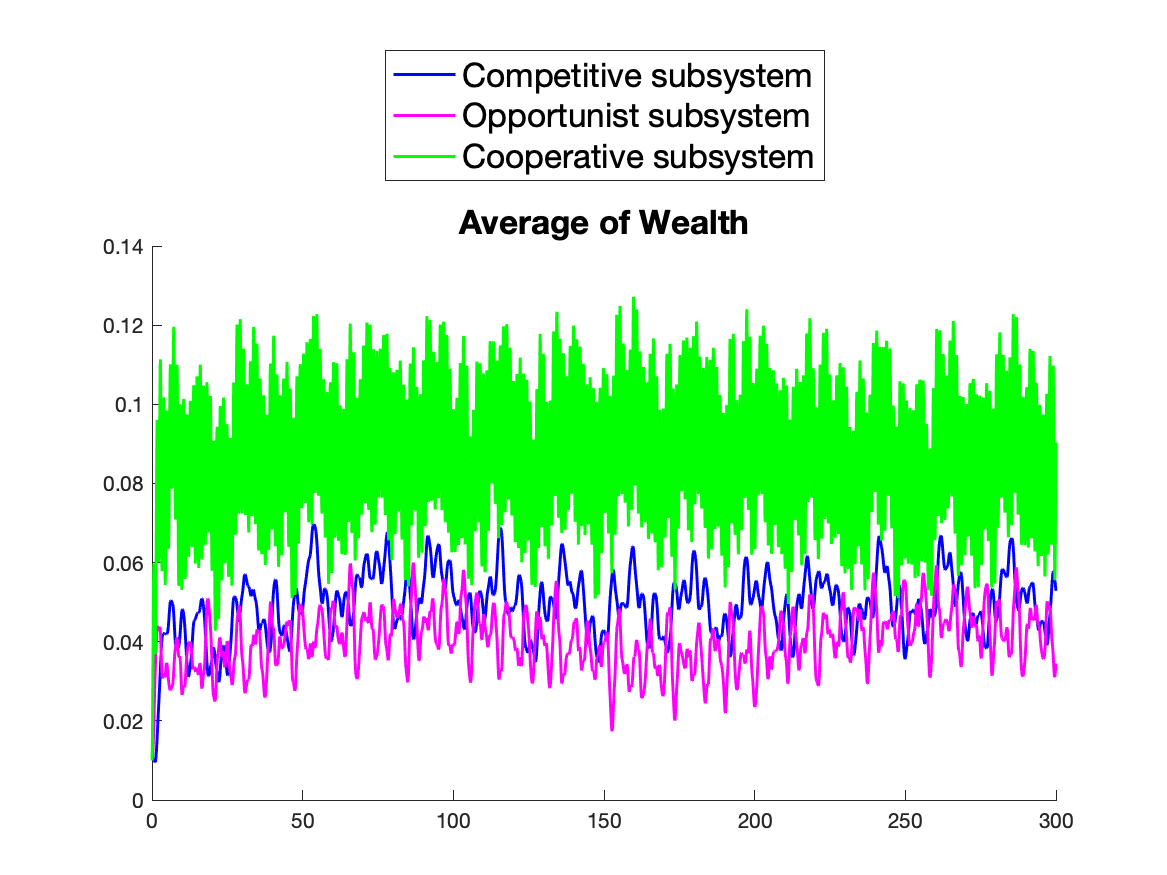}}
\subfigure[]{\includegraphics[width=0.47\textwidth]{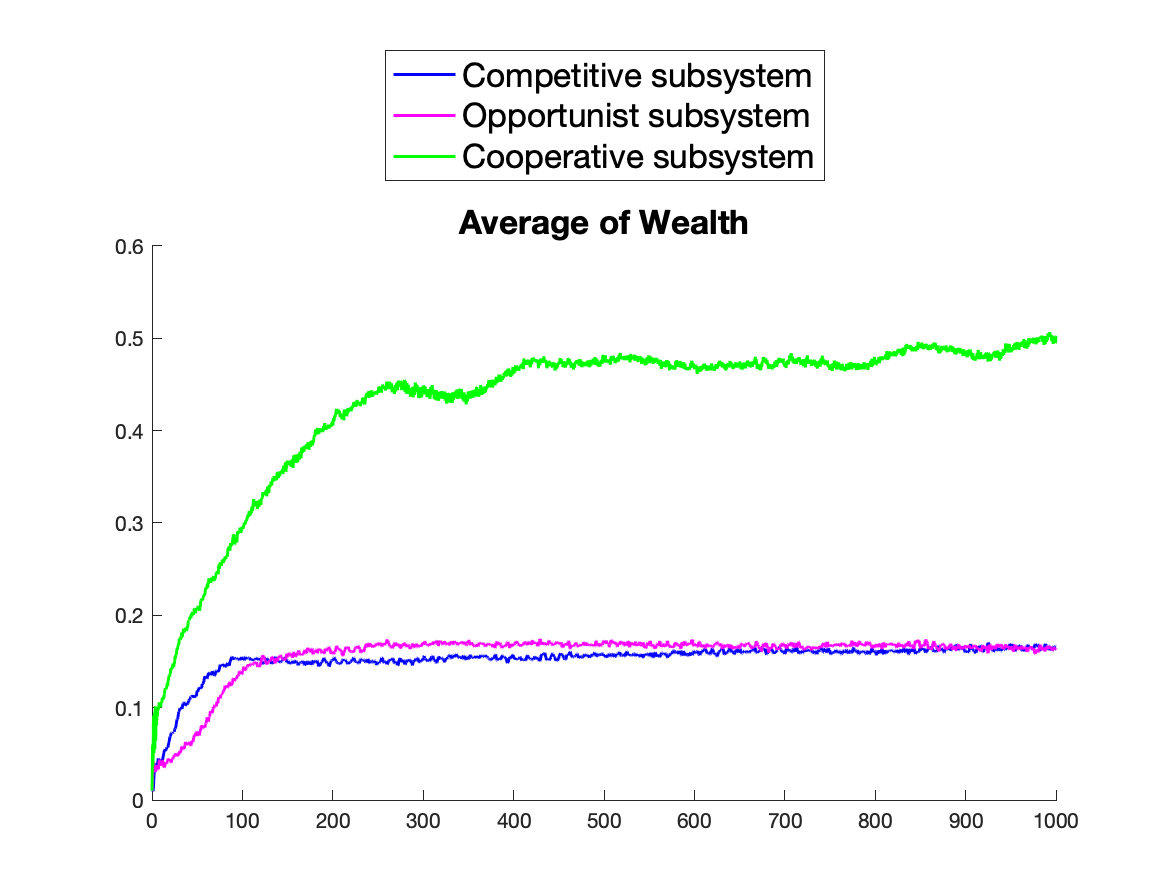}}\\
\subfigure[]{\includegraphics[width=0.47\textwidth]{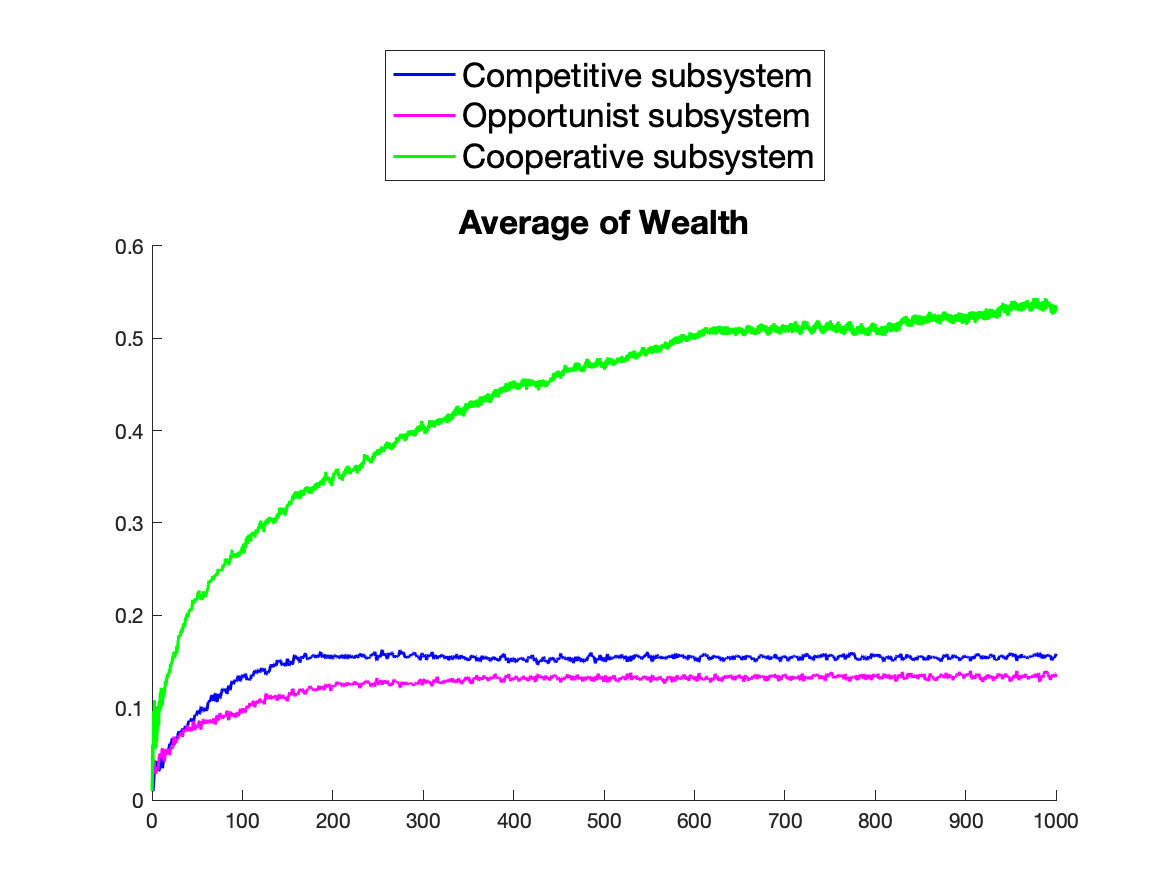}}
\subfigure[]{\includegraphics[width=0.47\textwidth]{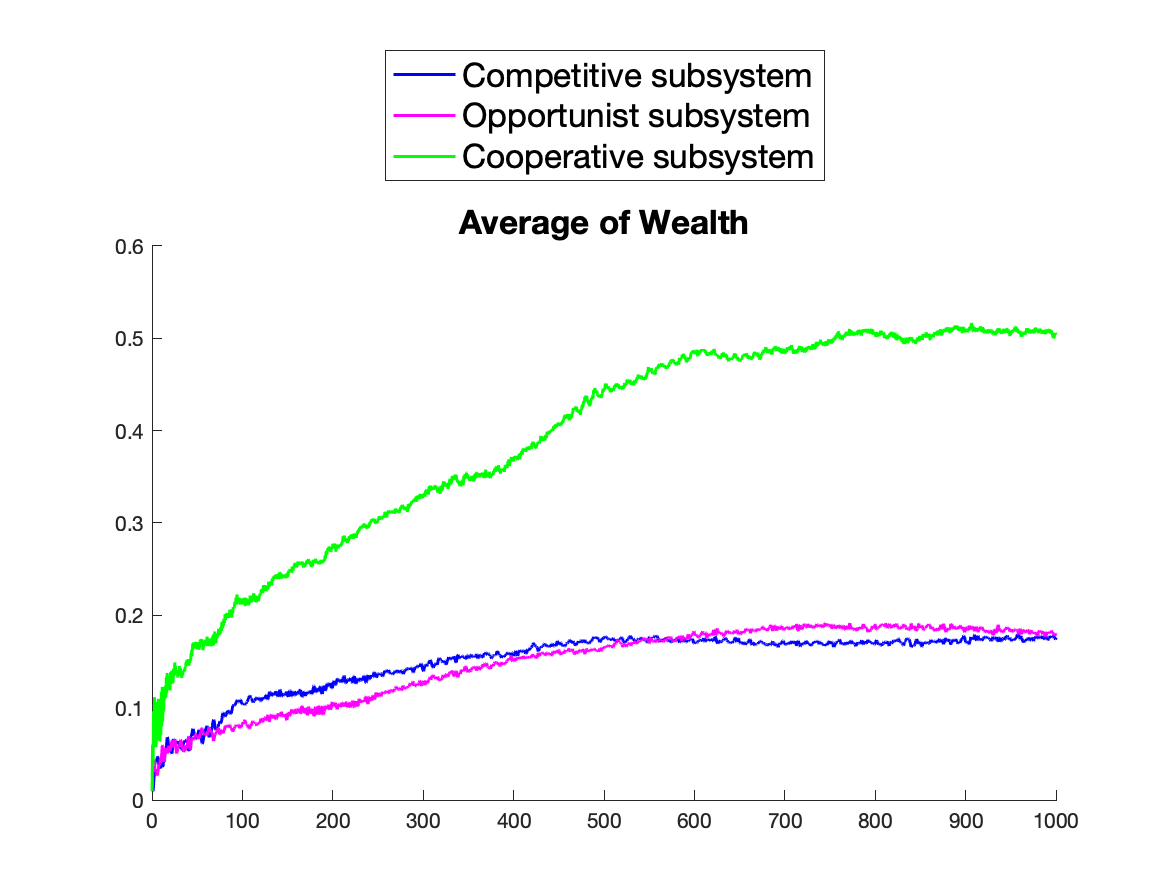}}
\caption{\label{fig:wealth100-34}Time evolution of wealth of the system with 100 agents: the 
cooperative and competitive subgroups contain 33 agents and the opportunist 
subgroup 34 agents; subfigure (a) displays the results without using the rule, subfigures (b), (c) 
and (d) the results using the $(\mathcal{H},\rho)$--induced dynamics approach with $\tau=1$, 
$\tau=2$ and $\tau=4$, respectively.}
\end{center}
\end{figure}

\begin{figure}
\begin{center}
\subfigure[]{\includegraphics[width=0.47\textwidth]{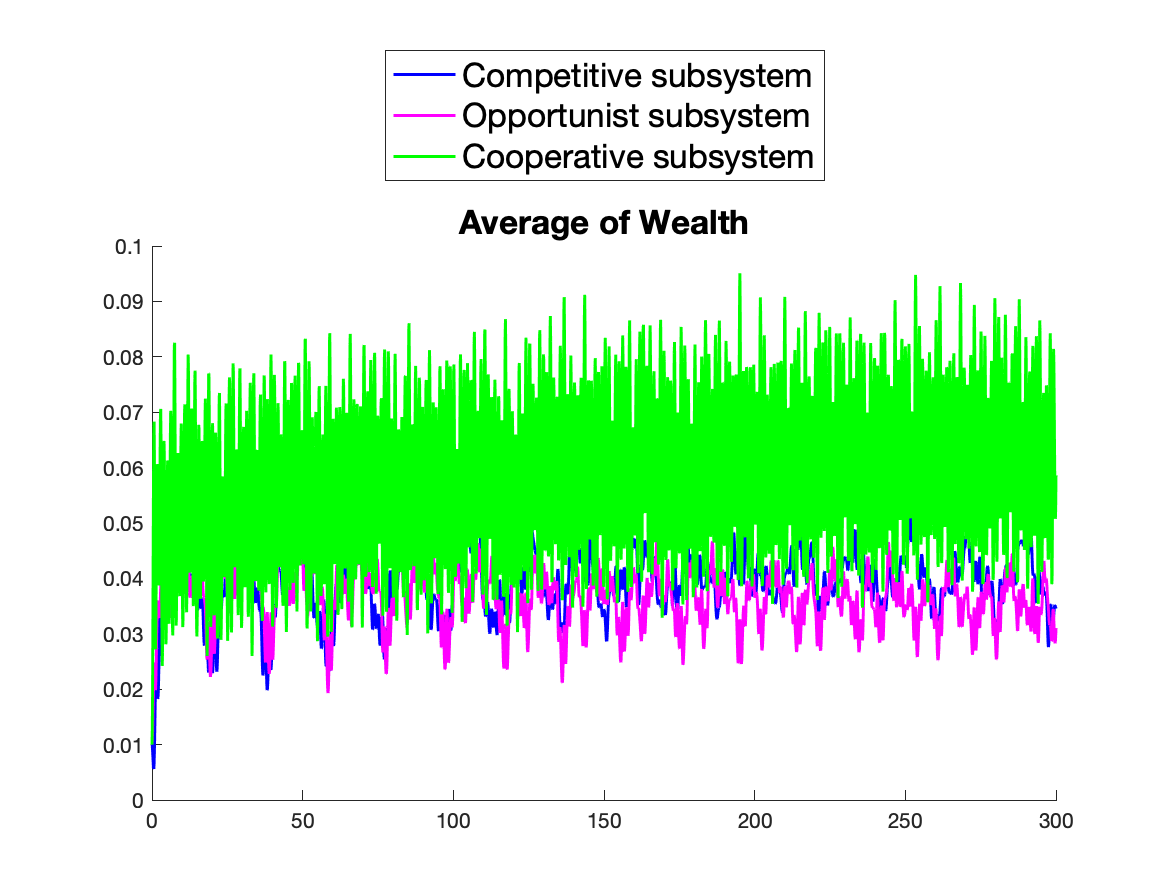}}
\subfigure[]{\includegraphics[width=0.47\textwidth]{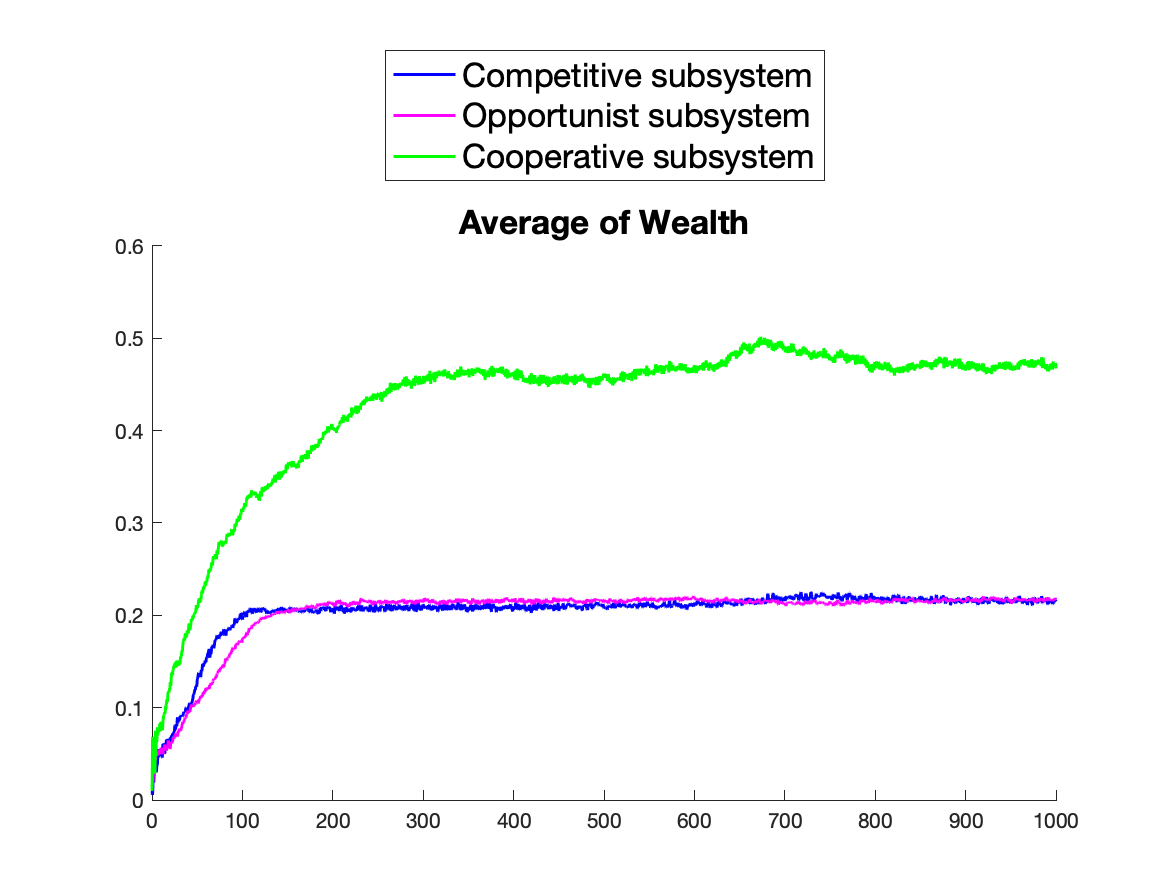}}\\
\subfigure[]{\includegraphics[width=0.47\textwidth]{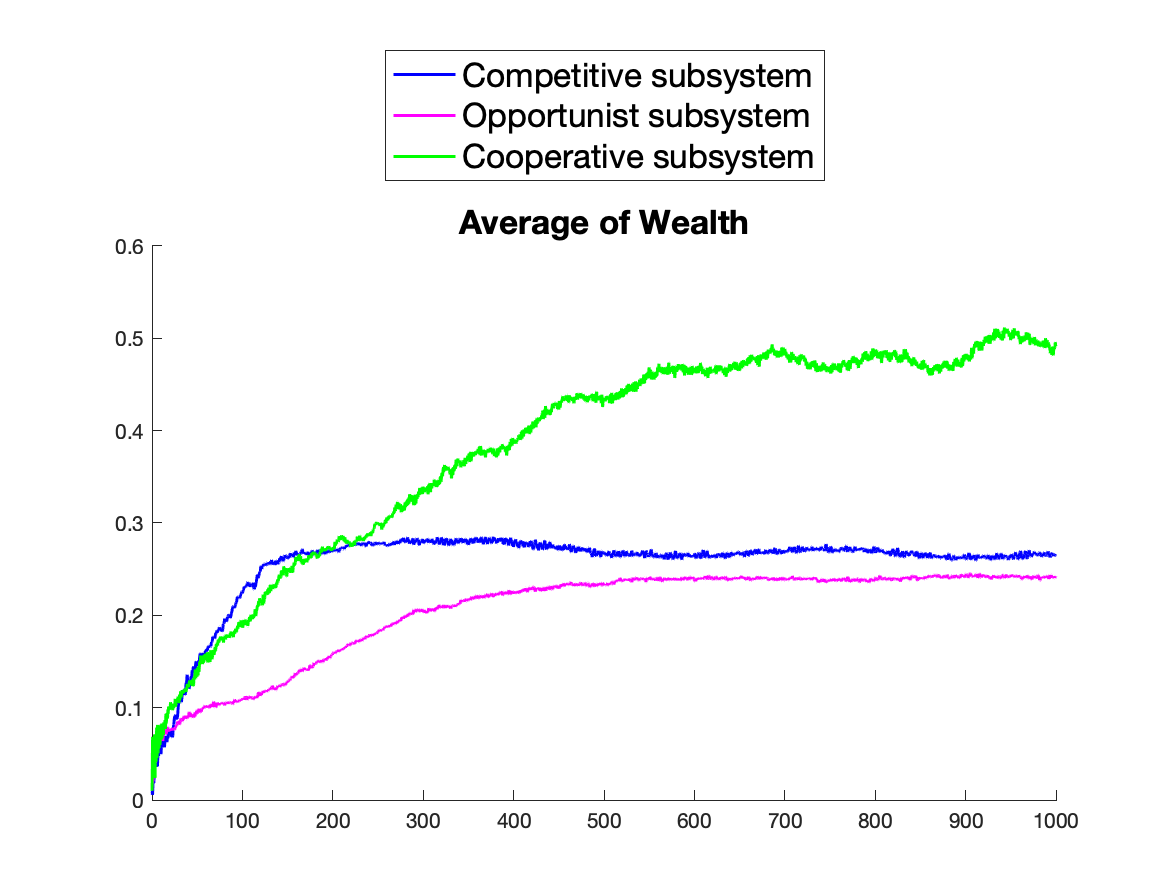}}
\subfigure[]{\includegraphics[width=0.47\textwidth]{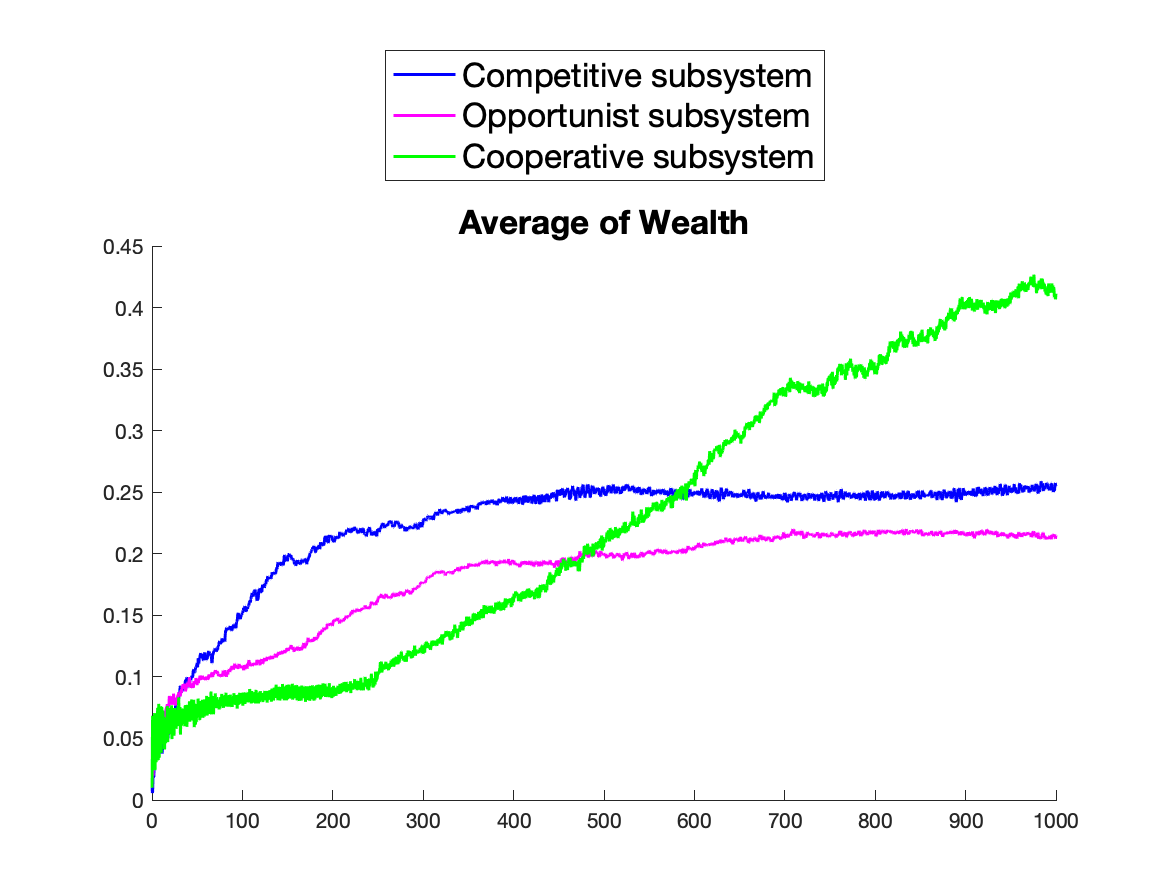}}
\caption{\label{fig:wealth100-50}Time evolution of wealth of the system with 100 agents: the 
cooperative and competitive subgroups contain 25 agents and the opportunist 
subgroup 50 agents; subfigure (a) displays the results without using the rule, subfigures (b), (c) 
and (d) the results using the $(\mathcal{H},\rho)$--induced dynamics approach with $\tau=1$, 
$\tau=2$ and $\tau=4$, respectively.}
\end{center}
\end{figure}

Let us choose randomly: 
\begin{itemize}
\item $N_1$ agents (the competitive subgroup) interacting each other with a competitive mechanism;
\item $N_2$ agents  (the cooperative subgroup) interacting each other with a cooperative 
mechanism; 
\item $N_3=N-N_1-N_2$ opportunist agents, \emph{i.e.}, each opportunist agent has a competitive 
interaction with an agent of the competitive subgroup, and a cooperative interaction with an agent of 
the cooperative subgroup; moreover, the opportunist agents compete each other.
\end{itemize}

We assume that when two agents are interacting in some way, in the case of competition, the 
parameter $\lambda_{j,k}$ is not constant but decreases with the distance $d(j,k)$ between the cells $j$ and $k$; 
on the contrary, in the case of cooperation, the parameter $\mu_{j,k}$ increases with $d(j,k)$. 
The values of these coefficients are chosen according to the relations
\[
\begin{aligned}
&\lambda_{j,k}=\lambda\left(1-\tanh\left(d(j,k)-d_{\hbox{max}}/2\right)\right),\\
&\mu_{j,k}=\mu\left(1+\tanh\left(d(j,k)-d_{\hbox{max}}/2\right)\right),
\end{aligned}
\] 
where $\lambda$ and $\mu$ are constants, that we choose both equal to $0.1$.

Finally, as far as the inertia parameters are concerned, each agent has an inertia parameter randomly chosen 
in the range between $0.5$ and $0.7$.

The numerical integration of the dynamics equations, using the standard Heisenberg view and 
the $(\mathcal{H},\rho)$--induced dynamics approach, provide some interesting results. In all the 
simulations all the agents start with the same initial amount of wealth, say $1/N$; moreover, the 
number of the purely competitive agents ($N_1$) is equal to the number of the purely cooperative 
agents ($N_2$); five different values of  $N_3$ (the number of opportunist agents) are 
considered.

To fix the rule,  let us define
\[
\begin{aligned}
& \delta_j^{(k)}=n_j(k\tau)-n_j((k-1)\tau), \quad k\in\mathbb{N},\\
& \delta^{(k)}=\hbox{max}\left\{ \left\vert \delta_j^{(k)} \right\vert ,\;  j=1,\ldots,N\right\};
\end{aligned} 
\]
then, at times $k\tau$ the inertia parameters change according to the law:
\[
\omega_j=\omega_j\left(1+\frac{\delta_j^{(k)}}{\delta^{(k)}}\right).
\]

Various simulations are carried out with increasing values of the number of opportunist agents.
Figures~\ref{fig:wealth100-10}, \ref{fig:wealth100-16}, \ref{fig:wealth100-20}, \ref{fig:wealth100-34} and \ref{fig:wealth100-50} (the opportunist subsystem made by 10, 16, 20, 34 and 50 agents, respectively) show the average wealth of the three subgroups of agents in the various cases,  that is, using the classical Heisenberg picture, and adopting the $(\mathcal{H},\rho)$--induced  dynamics approach with $\tau=1,2,4$.
Without the rule, the evolution is trivial and a never ending oscillatory outcome is obtained. On the contrary, we can observe that, in general, using the $(\mathcal{H},\rho)$--induced dynamics approach, cooperation gives better results in terms of wealth; cooperative subgroup, as time increases, obtains an amount of average wealth always greater than that of the purely competitive subgroup and often greater than that of the opportunist subgroup. For a low number of opportunist agents (Figure~\ref{fig:wealth100-10}), the time needed for having an average wealth of the cooperative subsystem greater than that of the opportunist subsystem increases with $\tau$: in some sense, a lower frequency of adjusting the attitudes of the agents makes a better performance for the opportunist agents for longer times. When the number of opportunist agents increases, their average wealth definitely becomes less than that of the cooperative subsystem, regardless the choice of $\tau$; also, the average wealth of opportunist subsystem is similar to that of competitive subsystem. This could be interpreted saying that in a system made by interacting agents an opportunist behavior may be for some time successful if the number of opportunist agents is not too high\footnote{``You can fool all the people some of the time, and some of the people all the time, but you cannot fool all the people all the time'', Abraham Lincoln.}. 
We are conscious that these results are only preliminary and that a deeper investigation with systems involving a large number of agents in different scenarios is needed, at least if we want to compare the performances of the behaviors of the agents (cooperative, competitive or opportunist).
 
\section{Concluding remarks}
\label{sec:conclusions}

In this paper, we presented a fermionic operatorial model describing a system where some agents 
interact each other both with a competitive and a cooperative mechanism. The dynamics is governed 
by a self--adjoint time--independent quadratic Hamiltonian operator. To enrich the dynamics, which 
can be at most quasiperiodic, we used the approach of $(\mathcal{H},\rho)$--induced dynamics
\cite{BDSGO-PhysicaA}. 
This variant of classical Heisenberg dynamics, modifying the inertia parameters of the agents according 
to the evolution of the system itself,  provided more interesting dynamical outcomes. The mean 
values of the number operators associated to the agents of the system represent a measure 
of their wealth. Using the $(\mathcal{H},\rho)$--induced dynamics, it is observed that the 
cooperating agents definitively increase their wealth better than the competitive and the 
opportunist agents. Moreover, a sort of irreversibility in the time evolution of wealth emerges.
This behavior is observed both in a toy model with seven agents and in a more sophisticated model where the 
agents are distributed on a one--dimensional torus, and the interaction parameters between two agents depend on the
distance. 

What the simulations seem to show is that the $(\mathcal{H},\rho)$--induced dynamics approach, 
taking into account the change of attitudes of the agents as a consequence of the evolution, 
produced an outcome where cooperative agents
tend to achieve a wealth greater than that reached by competitive and opportunist agents. 

Work is in progress with the aim of considering a model with a large number of agents who, 
according to the evolution of their states, are also able to change their interaction style 
(switching from competition to cooperation, and vice versa) as well as the values of the 
interaction parameters.

\section*{Acknowledgements}
This work has been carried out with the patronage of ``Gruppo Nazionale per la Fisica Matematica'' (GNFM) 
of ``Istituto Nazionale di Alta Matematica'' (INdAM). 

\section*{Funding}
M.G. acknowledges financial support from \textit{Finanziamento del Programma Operativo Nazionale (PON) ``Ricerca e Innovazione'' 2014-2020 a valere sul\-l'Asse IV ``Istruzione e ricerca per il recupero''- Azione IV-Dottorati e contratti di ricerca su tematiche dell’innovazione, CUP J45F21001750007}.

\end{document}